\documentclass[journal,letterpaper,twoside]{IEEEtran}
\usepackage[subpreambles=true]{standalone}
\usepackage{microtype}
\usepackage{graphicx}
\usepackage{booktabs} % for professional tables
\usepackage{amsmath,amsfonts,amsthm,bm} % Math packages
\usepackage{tikz}
\usetikzlibrary{bayesnet}
\usetikzlibrary{fit,positioning}
\usepackage{accents}
\usepackage{commath}
\usepackage[sc,osf]{mathpazo}
\usepackage{caption}
\usepackage{subcaption}
\usepackage{float}
\restylefloat{table}
\usepackage{algorithmic, algorithm}
\usepackage{authblk}

\usepackage{url}
\title{A Graphical Model for Fusing Diverse Microbiome Data}

\author[1,*]{Mehmet Aktukmak}
\author[1]{Haonan Zhu}
\author[2,3]{Marc G. Chevrette}
\author[3]{Julia Nepper}
\author[3,4]{Shruthi Magesh}
\author[2,3]{Jo Handelsman}
\author[1,*]{Alfred Hero}
\affil[1]{Department of Electrical and Computer Engineering, University of Michigan}
\affil[2]{Department of Plant Pathology, University of Wisconsin-Madison}
\affil[3]{Wisconsin Institute for Discovery}
\affil[4]{Microbiology Doctoral Training Program, University of Wisconsin-Madison}
\affil[*]{Corresponding Authors: aktukmak@umich.edu, hero@eecs.umich.edu}

\begin{document}

\maketitle

\begin{abstract}
     This paper develops a Bayesian graphical model for fusing disparate types of count data. The motivating application is the study of bacterial  communities from diverse high dimensional features, in this case transcripts, collected from different treatments. In such datasets, there are no explicit correspondences between the communities and each correspond to different factors, %types of species genomes,
     making data fusion challenging.
     We introduce a flexible multinomial-Gaussian generative model for jointly modeling such count data. This latent variable model jointly characterizes the observed data through a common multivariate Gaussian latent space that parameterizes the set of multinomial probabilities of the transcriptome counts. The covariance matrix of the latent variables induces a covariance matrix of co-dependencies between all the transcripts, effectively fusing multiple data sources. We present a computationally scalable variational Expectation-Maximization (EM) algorithm for inferring the latent variables and the parameters of the model. The inferred latent variables provide a common dimensionality reduction for visualizing the data and the inferred parameters provide a predictive posterior distribution. In addition to simulation studies that demonstrate the variational EM procedure, we apply our model to a bacterial microbiome dataset.\footnote{This work was partially supported by grants from ARO W911NF-19-102 and DOE DE-NA0003921. }
\end{abstract}

\section{Introduction}

We introduce a Bayesian graphical model for jointly modeling and fusing high dimensional count data collected from different sensors with no explicit correspondences between their feature sets. Our model is relevant to the many areas of multi-modality fusion where data is collected from diverse but incommensurate sensor modalities. Examples include multi-view learning in computer vision and automated language translation in natural language processing. However, this paper focuses on a particularly timely application: fusion of microbiome data from diverse microbial communities.

The analysis of meta-transcriptomic data has been of increasing interest to researchers in the biological and health sciences. Microbiomes exist in diverse environments and are critical to sustaining life, balancing ecosystems, and producing antibiotics, among many other functions.  Microbiomes consist of communities of microbes that interact with each other to maintain stability and resilience to environmental conditions and microbial intrusions from competitors. It has therefore been of great scientific interest to quantify changes in microbiome communities due to changing conditions using experimental data. For example, one area of study is the rhizosphere, which is a community of microbial species living around plant root systems, known to be sensitive to environmental factors \cite{fitzpatrick2020plant}. Another area of study is the spectrum of responses of microbiomes to stressors, collectively called the microbial exposome \cite{miller2013exposome}.  

One of the principal sensing platforms used to study microbiome communities applies gene sequencing to a microbiome sample, e.g., collected from the gut, the soil, or other environment. 
A common way to obtain a global profile of a microbial community is to perform gene sequencing on a biological sample.  For example, RNA-Seq % or meta-transcriptomics, which 
measures gene expression in a community by quantifying the number of times each gene transcript occurs in the pool of sequenced RNAs. Each microbial species in the community is represented by its own unique set of transcripts, i.e., its transcriptome, and fusing information from different transcriptomes yields the global profile of gene expression across all species in the community. This type of analysis is known as metatranscriptomics and it provides a functional profile of the community that can complement the gene taxonomic profiling provided by metagenomics \cite{rohlf1981comparing,mau1997phylogenetic,aguiar2016metagenomics}.  This paper introduces a Bayesian graphical model for the metatranscriptomics problem, where inference is performed using a scalable variational EM inference method.  Notably, our model can capture patterns of similarity between histograms of different species' gene expression without inter-species genome-to-genome mappings or knowledge of inter-species transcriptomic pathway correspondences.

% Motivation
A main feature of our model is that it estimates the global covariance structure of gene expression when the observations are in the form of count vectors produced by RNA-Seq. Correlations between transcript abundances are informative about the effect of environmental conditions on microbial communities % of abiotic factors, environmental conditions and indirect associations  
\cite{shakya2019advances}. 
%Specifically, we are interested in modeling the density of abundance datasets associated with multiple microbial communities. 
%The samples are collected from several sites in different environments. 
In particular, the global covariance matrix captures inter- and intra species interactions. For example, the expression of a single gene in a species can influence other gene expressions in that species or the gene expressions of other community members. %There may also be also unmeasured environmental conditions or mediator species, which may effect gene expression throughout the community, changing the covariance structure.  % \cite{blachet2020} it is
% besides the interactions of the taxons of interest. 
%Consequently, the distribution of the taxons may result from different, possibly, unknown factors and interactions.
We propose a latent variable graphical model that can capture the hidden factors underlying such dependencies.
%between observed count vectors associated with the taxons of multiple communities. %Although the model represented here heavily focus on abundance data analysis in terms of terminology, it can be applied to any datasets exhibiting count vectors, such as text and document collections from multiple languages. 
    
The main assumption underlying our proposed model is the existence of a hidden low-dimensional continuous latent space that can explain the observed data. We model the observations as conditionally multinomial distributed given the latent variables, which are assumed to be multivariate Gaussian with low rank covariance structure. % to be inferred from the data. 
%captures correlations of the observed variables are then captured in the latent space through the covariance parameter of the Gaussian distribution. However, 
Due to the lack of conjugacy between the Gaussian and multinomial distributions, exact Bayes inference is not tractable. We therefore adopt a Bayes variational inference approach \cite{bottou2018optimization, blei2017variational} to develop an algorithm for estimating the parameters of the proposed model and projecting the data to the latent space.  
%that can infer the latent variables as well as model parameters in a computationally efficient way. 
    
    % Related models
    % LVM
The proposed model can be contrasted with previously introduced latent variable models used in multi-view learning and dimensionality reduction. Factor analysis (FA) \cite{pml2Book} is a classical method that is a  generalization of Principal Component Analysis (PCA) \cite{bishop2006pattern} and Probabilistic PCA \cite{tipping1999probabilistic}. FA decomposes the observed data matrix into a low dimensional set of factor loadings and factor scores, imposing a low-rank constraint on the covariance matrix. Like our proposed model, the FA model also assumes a low-dimensional Gaussian latent space but it does not account for the counting nature of the observed data.  
%{\color{red} We should review and reference other FA models that are derived under other observation distributions. For example: Taddy, Matt. "Multinomial inverse regression for text analysis." Journal of the American Statistical Association 108, no. 503 (2013): 755-770.}%only deal with continuous observations due to Gaussian observation model. 

% Multinomial data
Several latent variable models have been proposed for counting observations. These include Latent Semantic Analysis \cite{landauer1997solution}, Multinomial PCA \cite{buntine2005discrete}, and Latent Dirichlet Allocation (LDA) \cite{blei2003latent}. LDA is the most closely related model to the model proposed here since 
%which is the most popular among the aforementioned algorithms, 
it is also a Bayesian graphical model for count data and uses multinomial distribution. The main difference is that LDA uses a Dirichlet distributed latent space instead of a Gaussian distributed latent space. Our Gaussian distributed latent space 
%is assumed to be Dirichlet distributed, in which it is not straightforward to capture a 
makes it possible to recover a non-trivial covariance structure among the count variables, unlike LDA \cite{blei2006correlated, murphy2012machine}. 
% Hierarchical modeling
% Recent models
% GGMs- Copulas and Microbial studies
    
Another way to capture the covariance structure of the observed variables is to ignore the counting nature of the data and use Gaussian Markov random fields (GMRF)  \cite{harris2016inferring} to directly estimate the covariance, or Gaussian Graphical Models (GGM) \cite{mazumder2012graphical} to enforce sparsity on the inverse of the covariance estimate. 
%which is quite popular in microbial data analysis. 
%The precision matrix is usually constrained to be sparse. Although this model can capture the covariance, 
Unlike our proposed multinomial-Gaussian model the GMRF and GGM do not incorporate a low rank latent structure on the covariance nor do they account for the counting nature of the data. 
%limited in terms of obtaining low-dimensional data representation. 
There have been extensions of the GGM to handle multinomial observations using copulas \cite{liu2009nonparanormal} that have been applied to microbiome analysis \cite{popovic2018general, popovic2019untangling,yoon2019microbial}.
    
    % Variational inference
Inference in latent variable models, like the one we propose here, can be challenging.  
%On the inference aspect, fitting a latent variable model is usually non-trivial, 
This is especially difficult when there is lack of conjugacy between the distributions of the latent variables and the observed variables. One approach is to perform point estimation for both the latent variables and the parameters in an alternating fashion \cite{collins2001generalization}, but this is prone to over-fitting \cite{welling2008deterministic} and convergence issues. Another approach is to use Markov Chain Monte Carlo (MCMC) methods, which can be computationally expensive \cite{mohamed2008bayesian}, especially in high dimension. As an alternative, variational Bayes inference has shown much promise \cite{blei2017variational}. Note that
Variational Bayes 
%Although variational inference is fast and provides competitive result in terms of accuracy, model specific derivations are required to establish the inference algorithm
is not a general purpose method and must be tailored to the specific statistical model \cite{kucukelbir2017automatic}. 
%especially, when there is lack of semi-conjugacy, which is the case when Gaussian-Multinomial random variables are connected in the graphical model. 
% Approximations
When there is a lack of conjugacy, as is the case for the multinomial-Gaussian model in this paper, local variational bound approximations are often adopted \cite{bottou2018optimization}. Additionally, when there is a problematic expression in the joint density, such as LogSumExp or LogGamma function, which may prevent the inference of the latent variables, surrogate optimization transfer based on Taylor series expansion can be applied to approximate the non-linear function either with linear \cite{blei2006correlated} or quadratic \cite{bohning1992multinomial, jaakkola2000bayesian, braun2010variational, khan2010variational, khan2012stick} functions. %Consequently, the inference becomes easier. 
We adopt such a local variational bound approach for deriving an inference algorithm for our proposed model. 
    
% Summary and Contributions
We summarize our contributions as follows. First, we propose a novel multinomial-Gaussian graphical model to fuse and capture the low rank covariance structure in counting data of disparate types. Our low-dimensional continuous latent space formulation provides dimensionality reduction that can be used for visualization of the count vectors on a common space. 
%to obtain homogeneous data representations for the discrete data, 
Second, we develop a novel and computationally scalable optimization algorithm based on variational inference to fit the proposed model, which exploits variational local bound approximations. Third, we validate and illustrate the model and its inference algorithm on a synthetic dataset and a real-world bacterial microbiome dataset \footnote{The code and the dataset are available at \url{https://github.com/maktukmak/microbiome-thor}.}. 
%The code and the dataset used in these experiments are available on []. 
    
\section{Proposed Model}

In this section, we formally define our proposed model and its corresponding variational inference algorithm. Lastly, we discuss computational complexity. %Since we focus on abundance datasets, our terminology is slightly biased. Particularly, we will use community and dataset interchangeably. Also, the count of taxons may refer as the observed variable. Furthermore, there may be mutually exclusive environmental conditions from which the samples are collected. We will model them by using discrete control variables.

\begin{figure}
    \small
	\centering
	%\resizebox{6.5cm}{6.5cm}{%
		\begin{tikzpicture}[
			roundnode/.style={circle, draw=green!60, fill=green!5, very thick, minimum size=7mm},
			squarednode/.style={rectangle, draw=black!60, very thick, minimum size=5mm},
			]
			% Define nodes
			\node[latent]       (z) {$\bm{z}_{k,i}$};
			\node[obs, above=2cm of z, xshift=0cm]        (x) {$\bm{x}_{kl,i}$};
			%\node[squarednode, below=1cm of z, xshift=-1.5cm]        (mu) {$\bm{\mu}_{k,i}$};
			\node[squarednode, left=2cm of z]        (S) {$\bm{\Sigma}_{k}$};
			\node[squarednode, left=2cm of x]    (Th) {$\Theta_{k,l}$};
			%\node[obs, below=2cm of z]        (u) {$\bm{u}_{k,i}$};
			\node[squarednode, below=2cm of z]        (mu) {$\bm{\mu}_{k}$};
			
			% Connect the nodes
			\edge {z} {x};
			%\edge {mu} {z};
			\edge {S} {z};
			\edge {Th} {x};
			%\edge {u} {z};
			\edge {mu} {z};
			
			% Plates
			{\tikzset{plate caption/.append style={above=5pt of #1.north}}
								\plate [inner sep=0.35cm, xshift=-0.0cm, yshift=0cm] {zxy} {(z)(x)} {$i = 1:I_k $} };
			{\tikzset{plate caption/.append style={below=5pt of #1.south west}}
				\plate {xTh} {(Th)(x)} {$l = 1:L $} };
			
			{\tikzset{plate caption/.append style={below=5pt of #1.south}}
				\plate {k} {(z)(mu)(x)(Th)(S)(zxy)(xTh)} {$k = 1:K $} };

		\end{tikzpicture}
	%}
	\caption{Graphical model representation of the proposed latent variable model. $\bm{x}_{kl,i}$ corresponds to the $i$th sample of community $l$ collected from environment $k$. The variables $\{\bm{x}_{kl,i}\}_{l=1:L}$ share a common low-dimensional latent variable $\bm{z}_{k,i}$ that captures the hidden causes of the observations. %The mean of the latent variable, $\bm{\mu}_{k,i}$, depends on the covariates $\bm{u}_{k,i}$, which ensures the covariance matrix $\bm{\Sigma}_{k}$ captures only the conditional dependencies by controlling the known variables of the environment $k$
	}
	\label{fig:graph}
\end{figure}
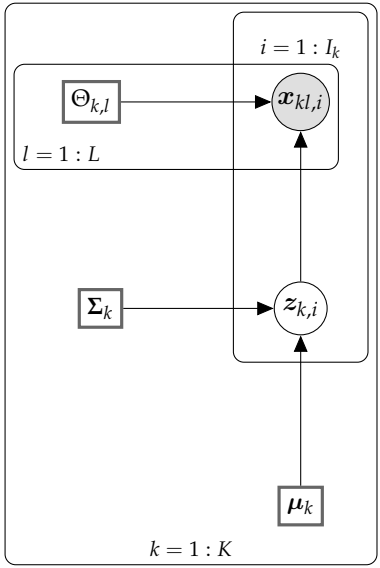

\subsection{Notation}

We denote the $i$th data sample of the $l$th species as $\bm{x}_{kl,i} \in \mathbb{Z}_+^{d_{l}}$, where $k$ indexes the experimental condition, and $d_{l}$ denotes the total number of transcripts for species $l$. The total number of experimental condition from which the samples are collected is denoted as $K$, and the total number of species in the model community is denoted as $L$, hence $l=1:L$ and $k=1:K$. For each experimental condition, different numbers of identically distributed samples are collected. Hence, we denote the total number of samples for the experimental condition $k$ as $I_k$. %If the environmental factors are measured for each sample, we denote those covariates as $\bm{u}_{k,i} \in \mathbb{R}^{D_{c}}$, where $D_{c}$ is the total number of features measured. 
The count vector for experimental condition $k$, species $l$, and replicate $i$ is denoted  $\bm{x}_{kl,i}$. The dataset for experimental condition $k$ is $D_{k} = \{ \{\bm{x}_{kl,i}\}_{l=1}^L \}_{i=1}^{I_k}$.

\subsection{Latent Variable Model} \label{sec:model}

We model the observed multi-species model community data as generated from a low dimensional latent variable generative model. Under this model the data are conditionally multinomial distributed given the latent variables, which are themselves Gaussian distributed with mean $\bm{\mu}_{k}$ and covariance matrix $\bm{\Sigma_k}$. As will be shown below, the model fuses the observed data across species and it induces a low rank decomposition of the population transcriptome covariance. 
%The proposed model is a generative latent variable model. Particularly, we assume lower dimensional latent variables to explain the observed count vectors, and we aim to capture the correlations between the variables via a low-rank covariance matrix. 
Let $\bm{z}_{k,i} \in \mathbb{R}^{d_z}$ be the latent variable assigned for the data sample $D_{k,i}$. $\bm{z}_{k,i}$ thus has the following multivariate normal prior distribution:

\begin{equation} \label{eq:z_pr}
	p(\bm{z}_{k,i} | \bm{\mu}_{k} , \Sigma_k) = \mathcal{N} (\bm{z}_{k,i}  | \bm{\mu}_{k} , \Sigma_k),
\end{equation}
where $\bm{\mu}_{k} \in \mathbb{R}^{d_z}$ is the prior mean vector and $\Sigma_k \in \mathbb{S}_{++}^{d_z}$ is the positive definite prior covariance matrix. The observed data consists of count vectors of the transcriptomes, which are modeled as multinomial distributed \cite{murphy2012machine}. We model the conditional distributions of the observed count vectors of species $l$ as follows:

\begin{equation} \label{eq:x_pr}
	p(\bm{x}_{kl,i} | \bm{z}_{k,i}, \Theta_{kl}) = Mu(\bm{x}_{kl,i} | N_{kl,i}, \mathcal{S}(\Theta_{kl}^T \bm{z}_{k,i}),
\end{equation}
where $Mu$ denotes the multinomial distribution, and $N_{kl,i}$ is the total number of counts of the $i$th data sample of species $l$. Note that we have introduced one more model parameter for each species, specifically $\Theta_{kl} \in \mathbb{R}^{d_{l} \times d_{z}}$, that maps lower dimensional latent space to the higher dimensional observation space of species $l$. Also note that both the latent variable $\bm{z}_{k,i}$ and the parameter $\Theta_{kl}$ are real-valued. Therefore, to provide a proper simplex support set for the multinomial distribution, we use the soft-max function, $\mathcal{S}(\bm{\eta})_d = \exp{\eta_d} / \sum_{d^{'}=1}^D \exp{\eta_{d^{'}}} $. The output of this function is a proper probability vector, i.e., $\sum_{d=1}^D \mathcal{S}(\bm{\eta})_d = 1$ and $\mathcal{S}(\bm{\eta})_d \ge 0$ for all $d = 1:D$. See Fig. \ref{fig:graph} for a graphical representation of the proposed model.

% Covariates
%In the case when the measurements contains site specific covariates, conditional joint density may be of more interest mainly due to need for capturing conditional dependencies between Taxons by controlling the known environmental variables. To model the conditional joint density, we make the prior mean vector as an affine function of the covariates, i.e., $\bm{\mu}_{k,i} = \bm{u}_{k,i} \bm{V}^T $. Note that when the covariates are observed, the prior mean vector is per sample, which is denoted as $\bm{\mu}_{i,k}$ with index $i$, otherwise a global parameter which is represented as $\bm{\mu}_{k}$. We estimate $\bm{V}$ when covariates exist, otherwise $\bm{\mu}_{k}$ (See Section \ref{sec:mstep}).

% Why gauss multinomial selected
Although it is natural to model the observed counts as multinomial distributed, it may not be obvious why we use  Gaussian latent variables for the latent space. A conjugate distribution such as Dirichlet may seem more natural than the Gaussian distribution, which is not conjugate to Multinomial. However, the components of the Dirichlet distribution are nearly independent \cite{blei2006correlated}, hence it is non-trivial to capture the correlations between the hidden components. On the other hand, the Multivariate normal distribution has a covariance parameter which specifically captures the correlation between the hidden components. This is useful for modeling the correlation between multiple datasets. Similar model assumptions are also adopted in topic models \cite{blei2006correlated, srivastava2017autoencoding}, categorical PCA \cite{khan2010variational}, and Gaussian process classification \cite{khan2012stick}. 
% Exclusive environments
Note that, although the communities are dependent through the latent variables, the experimental conditions are modeled as independent. Hence, there is no coupling between the experimental conditions and thus we fit independent  models for each condition.%, which ensures controlling the discrete environmental conditions. 

The joint log likelihood of the proposed model is of the form $\sum_{k=1}^{K}\sum_{i=1}^{I_k} \log p(\bm{z}_{k,i}, D_{k,i})$, where:
\begin{align}
    \begin{split}
	\log p(&\bm{z}_{k,i}, D_{k,i}) = \log p(\bm{z}_{k,i}) + \sum_{l=1}^L \log p(\bm{x}_{kl,i} | \bm{z}_{k,i}) \\
	= & -\frac{1}{2} \big[(\bm{z}_{k,i} - \bm{\mu}_{k})^T \Sigma_k^{-1} (\bm{z}_{k,i} - \bm{\mu}_{k}) + \log|\Sigma_k|\big]\\ 
	& + \sum_{l=1}^L\sum_{d=1}^D \bm{x}_{kl,id} ( \Theta_{kl,d} \bm{z}_{k,i} - \text{lse}(\Theta_{kl}  \bm{z}_{k,i}) ) + const,
	\end{split}
\end{align}
in which lse denotes the log-sum-exp function, i.e., log of the denominator of the soft-max function, and we suppress the deterministic parameters to avoid clutter. %in probability functions.
Taking the expectation with respect to $\bm{z}_{k,i}$ is tractable for the linear and quadratic terms, but intractable for the lse term.  We describe an asymptotic approximation in the next section.

\subsection{Optimization}

Next, we develop a variational EM maximum likelihood algorithm \cite{bottou2018optimization, blei2017variational} to infer the deterministic parameters $\bm{\mu}_{k}$, $\bm{\Sigma}_{k}$, and $\Theta_{kl}$. The main objective is to maximize the likelihood of the observations under the model. The algorithm comprises two alternating steps: i) Expectation step (E-step), where we integrate out the latent variables, ii) Maximization step (M-step), where we optimize the model parameters to maximize the marginal likelihood. 

\subsubsection{Objective}

The proposed model uses Gaussian latent variables for the multinomial observations. Due to lack of conjugacy between Gaussian and Multinomial distributions, the likelihood function is not closed form. Specifically, integrating out the latent variables becomes intractable (See Section \ref{sec:bound} for the details). Hence, we resort to variational inference, in which a lower bound on the likelihood function is derived and maximized. This lower bound is obtained by approximating the posterior distributions of the latent variables. In variational inference, the objective is to minimize the distance (KL-divergence) between the approximate and exact posterior distributions. This objective can be expressed for a single latent variable $\bm{z}_{k,i}$ as follows:

\begin{align}
    \begin{split}
	\mathbb{KL} (q_{\lambda_{k,i}} | p  ) &= \mathbb{E}_{q_{\lambda_{k,i}}} \log \big[ \frac{q(\bm{z}_{k,i} | \bm{\lambda}_{k,i})}{p(\bm{z}_{k,i} | D_{k,i})} \big] \\
	&= \mathbb{E}_{q_{\lambda_{k,i}}} \log \big[ \frac{q(\bm{z}_{k,i} | \bm{\lambda}_{k,i})}{p(\bm{z}_{k,i}, D_{k,i})} p(D_{k,i}) \big] \\
	&= \mathbb{E}_{q_{\lambda_{k,i}}} \big[ \log q(\bm{z}_{k,i} | \bm{\lambda}_{k,i}) - \log p(\bm{z}_{k,i}, D_{k,i}) \big] \\
	& \;\;\;\; + \log p(D_{k,i}),
	\end{split}
\end{align}
where $\bm{\lambda}_{k,i}$ corresponds to the set of parameters of the approximate posterior distribution $q(\bm{z}_{k,i} | \bm{\lambda}_{k,i})$. The expectation operator is defined as $\mathbb{E}_{q_{\lambda}}f(z) = \int f(z) q(\bm{z} | \bm{\lambda}) dz$. Note that the evidence (marginal likelihood) $p(D_{k,i})$ does not depend on $\bm{z}_{k,i}$. Hence, the negative of the expectation term forms a lower bound on the log evidence since the KL distance is always non-negative. This function is known as evidence lower bound (ELBO) and it is the objective function that is maximized in variational EM. The ELBO has the following form:
\begin{equation} \label{eq:obj}
	\mathcal{L} = \sum_{i=1}^I \sum_{k=1}^K \mathbb{E}_{q_{\lambda_{k,i}}} \big[ \log p(\bm{z}_{k,i}, D_{k,i}) - \log q(\bm{z}_{k,i} | \bm{\lambda}_{k,i}) \big],
\end{equation}
where the first term in the expectation corresponds to the joint distribution of the latent variable $\bm{z}_{k,i}$ and the associated observed data $D_{k,i}$. The second term corresponds to log of the approximate posterior distribution. The joint distribution has the following form:
\begin{equation} \label{eq:joint}
	\log p(\bm{z}_{k,i}, D_{k,i}) = \sum_{l=1}^L \log p(\bm{x}_{kl,i} | \bm{z}_{k,i}) + %\log p(\bm{z}_{k,i} | \bm{u}_{k,i}),
	\log p(\bm{x}_{kl,i} | \bm{z}_{k,i}).
\end{equation}
The expressions for $ p(\bm{x}_{kl,i} | \bm{z}_{k,i})$ and $p(\bm{x}_{kl,i} | \bm{z}_{k,i})$ are given in Eq. \ref{eq:x_pr} and Eq. \ref{eq:z_pr}, respectively. %The second term in Eq. \ref{eq:obj} is simply the log of the approximate posterior distribution. %The objective in Eq. \ref{eq:obj} requires taking expectations of the log joint distribution with respect to the approximate posterior distribution. 
We approximate the posterior distribution of $\bm{z}_{k,i}$ as Gaussian with the following form:
\begin{equation}
	q(\bm{z}_{k,i} | \bm{\lambda}_{k,i}) =  \mathcal{N}(\bm{z}_{k,i} | \bm{m}_{k,i}, \bm{S}_{k,i}),
\end{equation}
where $\bm{\lambda}_{k,i} = \{ \bm{m}_{k,i}, \bm{S}_{k,i} \}$ is the set of free parameters. Specifically, $\bm{m}_{k,i}$ is the posterior mean and $\bm{S}_{k,i}$ is the posterior covariance. The expectation of the approximate posterior distribution in Eq. \ref{eq:obj} corresponds to the Gaussian entropy function, which has a closed form expression. However, the expectation of the joint distribution is intractable to compute. Next, we present an approximation to resolve the issue.

\subsubsection{An upper bound on the LSE} \label{sec:bound}

To see why the conditional expectation is intractable, note that the explicit form of the log likelihood of $\bm{x}_{kl,i}$ is a multinomial distribution:

\begin{equation}  \label{eq:mult_ll}
	\log p(\bm{x}_{kl,i} | \bm{z}_{k,i}) = \sum_{d=1}^D \bm{x}_{kl,id} ( \Theta_{kl,d} \bm{z}_{k,i} - \text{lse}(\Theta_{kl} \bm{z}_{k,i}) ).
\end{equation}
Taking expectation corresponds to integrating out Gaussian distributed $\bm{z}_{k,i}$. The conditional expectation of the first term is easily determined since it linearly depends on $\bm{z}_{k,i}$. However, the expectation of the second term, which requires integrating $\bm{z}_{k,i}$ over the $\text{lse}$ function, is intractable to compute in a closed form. To overcome this issue, we perform quadratic surrogate optimization transfer, in which a quadratic approximation to the $\text{lse}$ function \cite{bohning1992multinomial} is applied. This results in an upper bound on the multinomial log likelihood. This approximation uses the second order Taylor series expansion with a fixed Hessian matrix. 
Particularly, the quadratic upper bound takes the following form:
\begin{equation} \label{eq:lse_app}
	\text{lse}(\Theta_{kl} \bm{z}_{k,i}) \le \frac{1}{2} \bm{z}_{k,i}^T \Theta_{kl}^T \bm{A}_{l} \Theta_{kl} \bm{z}_{k,i} - \bm{b}_{kl,i}^T \Theta_{kl} \bm{z}_{k,i} + \bm{c}_{kl,i},
\end{equation}
where 
\begin{equation} \label{eq:a}
    \bm{A}_{l} = 0.5 [ \bm{I}_{D_{xl}} - (1/(D_{xl}+1)) \bm{1}_{D_{xl}} \bm{1}_{D_{xl}}^T ]
\end{equation}
is a constant Hessian matrix, whose entries depend only on the dimension of the observation space. The other intermediate parameters $\bm{b}_{kl,i}$ and $\bm{c}_{kl,i}$ are given as follows:
\begin{equation}
	\bm{b}_{kl,i}= \bm{A}_{l} \Phi_{kl,i} - \mathcal{S}(\Phi_{kl,i}),
\end{equation}
\begin{equation}
	\bm{c}_{kl,i} = \frac{1}{2} \Phi_{kl,i}^T \bm{A}_{l} \Phi_{kl,i} - \mathcal{S}(\Phi_{kl,i})^T \Phi_{kl,i} + \text{lse}(\Phi_{kl,i}),
\end{equation}
where $\Phi_{kl,i}$ is the Taylor series expansion point, which is optimized as a free variational parameter. Note that intermediate parameters are deterministic function of $\Phi_{kl,i}$. Plugging the approximation in Eq. \ref{eq:lse_app} to Eq. \ref{eq:obj} results in a convex lower bound on ELBO, denoted as $\mathcal{L'}$, which is $\le \mathcal{L}$ and tight at $\Phi_{kl,i}$. Using $\mathcal{L'}$ resolves the intractable integration in Eq. \ref{eq:obj}, resulting in closed form posterior parameter estimates, as described in the next section.

\subsubsection{Posterior Distributions - E-step}

The E-step in the variational EM algorithm computes approximate posterior distributions of the latent variables, which are subsequently used to compute the expectations in Eq. \ref{eq:obj}. Particularly, there are two parameters to be estimated for each latent variable $\bm{z}_{k,i}$, which are the mean vector $\bm{m}_{k,i}$ and the covariance matrix $\bm{S}_{k,i}$. It is straightforward to maximize over these parameters by using the completing-the-square approach \cite{bishop2006pattern} (See Appendix \ref{ap:post}). The terms that quadratically depend on $\bm{z}_{k,i}$ in the joint log-likelihood yield the posterior covariance update:
\begin{equation} \label{eq:post_cov}
	\bm{S}_{k,i} = \big[ \Sigma_k^{-1}  +  \sum_{l=1}^{L} N_{kl,i} \Theta_{kl}^T \bm{A}_{l} \Theta_{kl} \big]^{-1},
\end{equation}
where $N_{kl,i}$ is the total number of counts of the $i$th data sample. Similarly, the terms that linearly depend on $\bm{z}_{k,i}$ yield the posterior mean update:
\begin{equation} \label{eq:post_mean}
	\bm{m}_{k,i} = \bm{S}_{k,i} \big[ \Sigma_k^{-1} \bm{\mu}_{k}  + 
	 \sum_{l=1}^{L} (\bm{x}_{kl,i} + N_{kl,i} \bm{b}_{kl,i}) \Theta_{kl}
	 \big].
\end{equation}
Lastly, we update the Taylor series expansion point as:
\begin{equation} \label{eq:post_taylor}
	\Phi_{kl,i} = \Theta_{kl} \bm{m}_{k,i}.
\end{equation}
Note that the update of $\Phi_{kl,i}$ depends on the posterior mean. Hence, the algorithm repeats the updates in Eq. \ref{eq:post_cov}, Eq. \ref{eq:post_mean}, and Eq. \ref{eq:post_taylor}, respectively, until convergence of the expansion point $\Phi_{kl,i}$.

\subsubsection{Point Estimates - M-step} \label{sec:mstep}

The M-step in the variational EM algorithm maximizes the ELBO with respect to the model parameters. Using the posterior distributions computed in the E-step, we compute the lower bound $\mathcal{L}^{'}$ by taking the expectations with respect to the posterior distributions. Afterwards, taking the derivatives with respect to the model parameters yields closed form update equations for the model parameters. Specifically, the updates for each $\Theta_{kl}$ are given as follows:
\begin{align} \label{eq:Theta}
    \begin{split}
	\Theta_{kl} = \Big[\sum_{i = 1}^{I_k} (\bm{x}_{kl,i} &+ N_{kl,i} \bm{b}_{kl,i}) \bm{A}_{l}^{-1} \bm{m}_{k,i} \Big] \\
	&\Big[ \sum_{i = 1}^I N_{kl,i} ( \bm{m}_{k,i} \bm{m}_{k,i}^T + \bm{S}_{k,i}) \Big]^{-1}.
	\end{split}
\end{align}
The update equations for the mean parameter and covariance of the prior distribution of $\bm{z}_{k,i}$ then follow as:
%\begin{equation}
%	\bm{V}_k = [\sum_{i=1}^{I_k} \bm{m}_{k,i}^T  \bm{u}_{k,i}] [ \sum_{i=1}^{I_k} %\bm{u}_{k,i} \bm{u}_{k,i}^T ]
%\end{equation}
\begin{equation} \label{eq:prior_mean}
	\bm{\mu}_k = \frac{1}{I_k} \sum_{i = 1}^I \bm{m}_{k,i},
\end{equation}
\begin{equation}\label{eq:prior_cov}
	\Sigma_k = \frac{1}{I} \sum_{i = 1}^I (\bm{m}_{k,i} - \bm{\mu}_k) (\bm{m}_{k,i} - \bm{\mu}_k)^T + \bm{S}_{k,i},
\end{equation}
respectively. The variational EM algorithm is summarized in Algorithm \ref{alg:opt}. 

Note that the model parameters of the proposed model are unidentifiable. Due to isotropic Gaussian prior on the latent variables, arbitrary rotation on $\Theta_{kl}$ results in same likelihood \cite{murphy2012machine}. This makes direct interpretation of the inferred latent variables ambiguous. Fortunately, this does not affect the predictive performance nor the predictor of covariance, which are the main focus of this paper.  %The prior mean vector is then $\bm{\mu}_{k,i} = \bm{V}^T \bm{u}_{k,i}$. In the case when no covariates exist, we don't estimate $\bm{V}$, hence the update takes the form $\bm{\mu}_k = 1 / I_k \sum_{i = 1}^I \bm{m}_{k,i} $.

\begin{algorithm}[H]
	\caption{Proposed Variational EM algorithm}
	\label{alg:opt}
	\begin{algorithmic}
	    \STATE {\bfseries Input:} $\{D_k\}_{k=1:K}$
    	\STATE Initialize $\{\bm{\mu}_k, \bm{\Sigma}_k, \{\bm{\Theta}_{kl}, \{\bm{\Phi}_{kl,i}\}_{i=1:I_k}\}_{l=1:L}\}_{k=1:K}$
	    \WHILE{not $\mathcal{L}^{'}$ converged}
    		\FOR{$k=1$ to $K$}
    			\FOR{$i=1$ {\bfseries to} $I_k$}
    				\STATE Infer posterior covariance $\bm{S}_{k,i}$ by Eq.
    				\ref{eq:post_cov} 
    				\STATE Infer posterior mean $\bm{m}_{k,i}$ by Eq.\ref{eq:post_mean}
    				\FOR{$l=1$ {\bfseries to} $L$}
    				    \STATE Update variational parameter $\bm{\Phi}_{kl,i}$ by Eq. \ref{eq:post_taylor}
    				\ENDFOR
    			\ENDFOR
    			\FOR{$l=1$ {\bfseries to} $L$}
        		    \STATE Estimate $\Theta_{kl}$ by Eq. \ref{eq:Theta}
        		\ENDFOR
        		\STATE Estimate $\bm{\mu}_k$  by Eq. \ref{eq:prior_mean}
        		\STATE Estimate $\bm{\Sigma}_k$ by Eq. \ref{eq:prior_cov}
        	\ENDFOR
    		\STATE Compute $\mathcal{L}^{'}$ by Eq. \ref{eq:obj}
    	\ENDWHILE
	\end{algorithmic}
\end{algorithm}

\subsection{Model-predicted Density}
\label{subsec: correlation analysis}
A predictor of the population covariance matrix can be extracted from the inferred model. There are two fundamental choices in the proposed model that pave the way to a predictor of the covariance matrix of the transcriptomes from the abundance data. First, we select Gaussian latent space that is common for all species, which models the observation covariance matrix with a low-rank decomposition due to inherent low-dimensional latent space. Second, we adopt a quadratic lower bound on the multinomial likelihood, hence the marginal likelihood of the observations can be approximated with a Gaussian distribution whose covariance matrix reveals the correlations between genomes. Particularly, we define a transformed version of the sample $\bm{x}_{kl,i}$ as $\tilde{\bm{x}}_{kl,i}$ with the following function:
\begin{equation}
	\tilde{\bm{x}}_{kl,i} = \bm{A}_{l}^{-1} ( \bm{b}_{kl,i} + \bm{x}_{kl,i} ),
\end{equation}
where $\bm{A}_{l}$ is the matrix defined in Eq. \ref{eq:a}. Then, it is straightforward to show that the likelihood of the transformed data $\tilde{\bm{x}}_{kl,i}$ is given as follow:
\begin{align} \label{eq:incuded_dist}
    \begin{split}
	p(\tilde{\bm{x}}_{kl,i}&|\Theta_{kl}, \bm{\mu}_k , \Sigma_k) \\ &= \int \mathcal{N} ( \tilde{\bm{x}}_{kl,i} | \Theta_{kl} \bm{z}_{k,i}, \bm{A}_{l}^{-1} ) \mathcal{N}(\bm{z}_{k,i} | \mathcal{N}(\bm{\mu}_k , \Sigma_k) d \bm{z}_{k,i} \\
	&= \mathcal{N} ( \tilde{\bm{x}}_{kl,i} | \Theta_{kl} \bm{\mu}_k, \bm{A}_{l}^{-1} + \Theta_{kl} \Sigma_k \Theta_{kl}^T ),
	\end{split}
\end{align}
where the covariance matrix $\bm{C}_{kl, \text{intra}} = \bm{A}_{l}^{-1} + \Theta_{kl} \Sigma_k \Theta_{kl}^T $ and the mean vector $\bm{\phi}_{kl} = \Theta_{kl} \bm{\mu_k}$ are of interest to us, in which $\bm{C}_{kl, \text{intra}}$ captures intra-species correlations of species $l$ in condition $k$. To obtain inter-species correlations, define $\tilde{\bm{A}}^{-1} = \text{diag}(\bm{A}_{1}^{-1}, \dots, \bm{A}_{L}^{-1})$ and $\tilde{\Theta}_{k} =  [ \Theta_{k1}, \dots, \Omega_{kL}]$, then  $\bm{C}_{k, \text{inter}} = \tilde{\bm{A}}^{-1} + \tilde{\Theta}_{k} \Sigma_k \tilde{\Theta}_{k}^T $ gives a covariance matrix for both inter-species and intra-species. To convert any covariance matrix to a proper correlation matrix, which is useful for visualization and analysis,  one can use the transformation $\text{Corr} = \text{diag}(\bm{C})^{-1/2} \bm{C} \text{diag}(\bm{C})^{-1/2} $. 

\subsection{Computational Complexity}

The computational complexity of the variational EM algorithm determines the algorithm's scalability to large datasets. For notational simplicity, we assume that there is only one discrete condition, hence we use $I$ instead of $I_k$. In the E-step, Eq. \ref{eq:post_cov} computes posterior covariance, which requires multiplication of a $d_z \times d_{l}$ matrix with its transpose resulting $O(d_z^2d_{l})$ complexity. This process is repeated for each species resulting in $O(L d_z^2 d_{l})$. Inverting the matrix for each sample costs $ O(Id_z^3)$. Hence, overall asymptotic complexity for the posterior covariance computation is $O(I ( d_z^3 + L d_z^2 d_{l}))$. 
The posterior mean computation in Eq. \ref{eq:post_mean} involves matrix-vector multiplications that require $O(L d_z d_{l})$, and $O(d_z^2)$ due to covariance posterior covariance multiplication. Hence, the total cost per sample is $O(L d_z d_{l} + d_z^2 )$ and the overall cost is $O(I (L d_z d_{l} + d_z^2))$. Consequently, the complexity of the E-step is $O(I (L d_z d_{l} + d_z^2 + d_z^3 + L d_z^2 d_{l}))$. Removing non-dominant terms results in $O(I (d_z^3 + L d_z^2 d_{l}))$. One can see that this scales linearly in terms of $L$, $d_{l}$, and $I$.
% M-step
On the other hand, the dominant computation in the M-step is for $\Theta_{kl}$. Eq. \ref{eq:Theta} comprises two terms. The first term requires $O(I d_{l} d_z)$ due to $I$ times vector-vector outer products. The second term requires $O(I d_z^2 + d_z^3)$ due to vector-vector outer products and subsequently matrix inversion. Multiplying these terms costs $O(d_{l} d_z^2)$, hence resulting total complexity of $O(L(I d_{l} d_z + I d_z^2 + d_z^3 + d_{l} d_z^2))$ for all $l=1:L$. It is also clear that this computation scales linearly in terms of $L$, $d_{l}$, and $I$. Modeling the conditions independently also induces linear complexity in terms of K. In summary, both E and M steps scale linearly in terms of $K$, $L$, $d_{l}$, and $I$, which suggests that the proposed optimization algorithm is scalable for large datasets as long as the latent space dimension $d_z$ is relatively small.

\begin{figure}[t]
    \centering
    \includegraphics[width=0.45\textwidth]{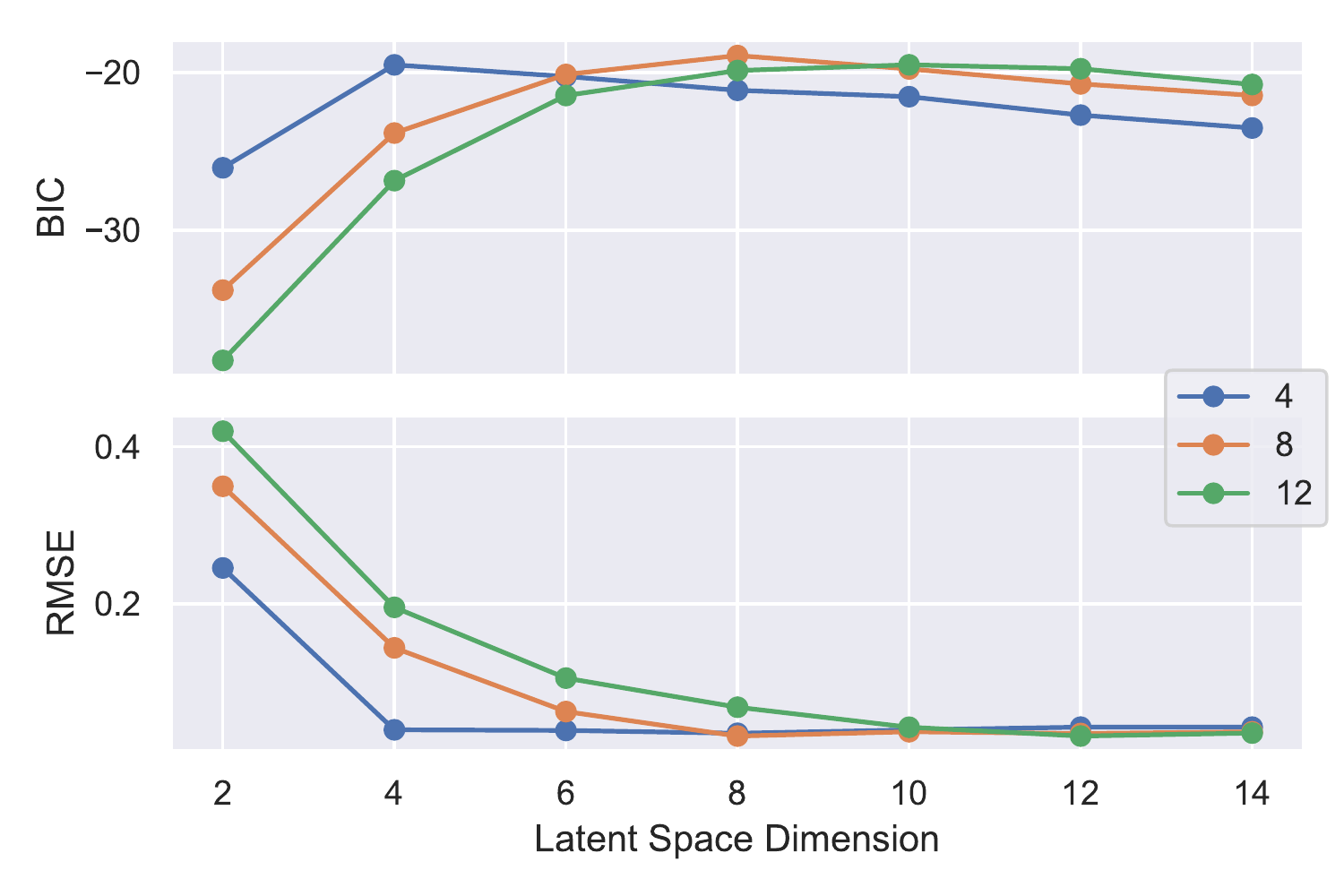}
    \caption{BIC approximation to the evidence and RMSE of the predicted covariance matrix with respect to the latent space dimension $d_z$. True dimensions are 4, 8, and 12. Blue, orange, and green curves show RMSE and the BIC penalized log likelihood (BIC), respectively. Note that the BIC exhibits a clear maximum over latent space dimension $d_z$. BIC values are scaled by factor $10^{-3}$.}
    \label{fig:rank}
\end{figure}

\begin{figure}[t]
    \centering
    \includegraphics[width=0.45\textwidth]{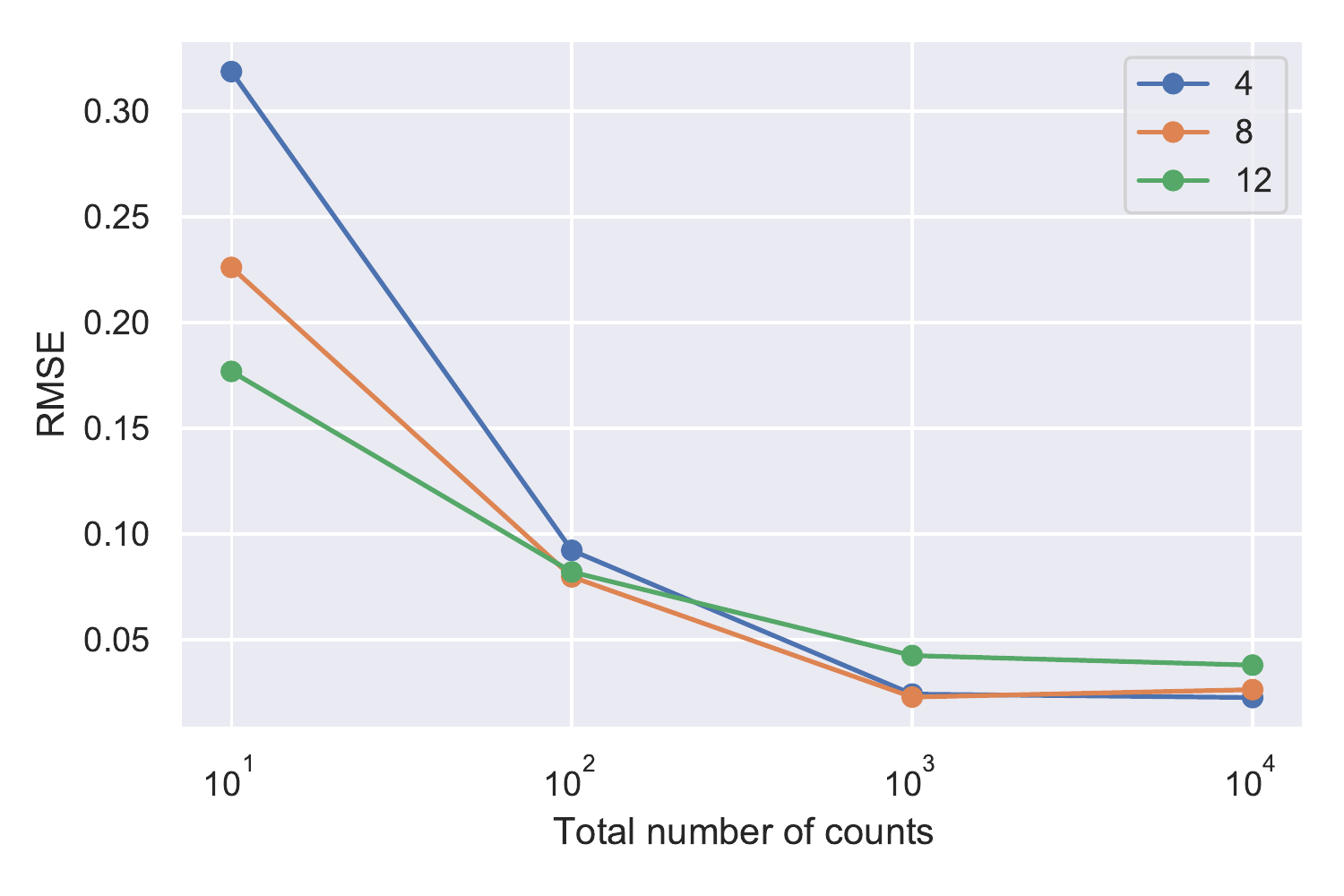}
    \caption{RSME of the covariance estimation with respect to average total number of counts observed in the metatranscriptomic data. As the counts increase the errors decrease until the counts reach a saturation limit.}
    \label{fig:counts}
\end{figure}

\begin{figure}[t]
    \centering
    \includegraphics[width=0.45\textwidth]{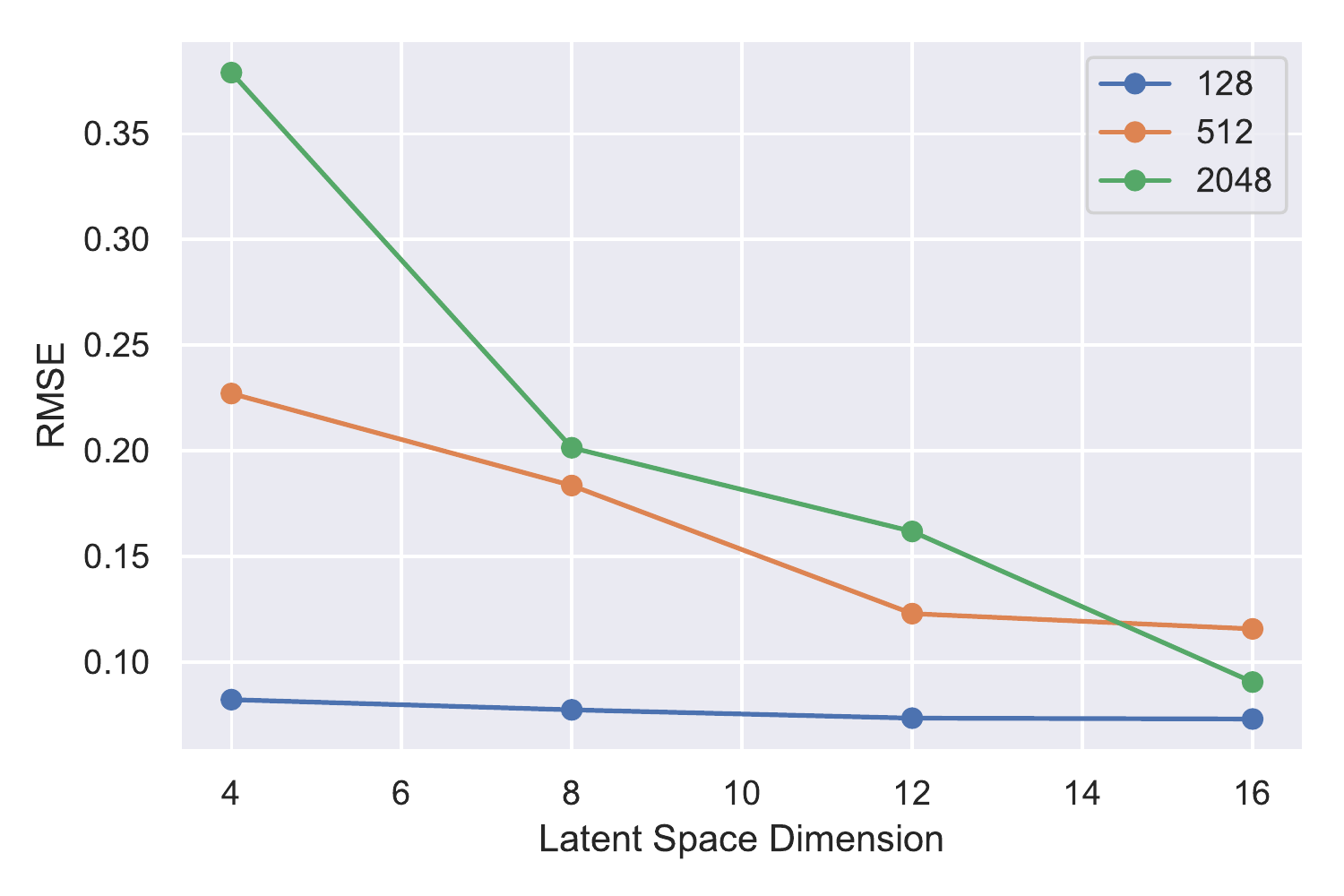}
    \caption{RMSE of the predicted covariance matrix with respect to the latent space dimension for three different observation space dimensions. }
    \label{fig:dims}
\end{figure}

\begin{figure}[t]
    \centering
    \begin{subfigure}[b]{0.45\textwidth}
        \centering
        \includegraphics[width=1\textwidth]{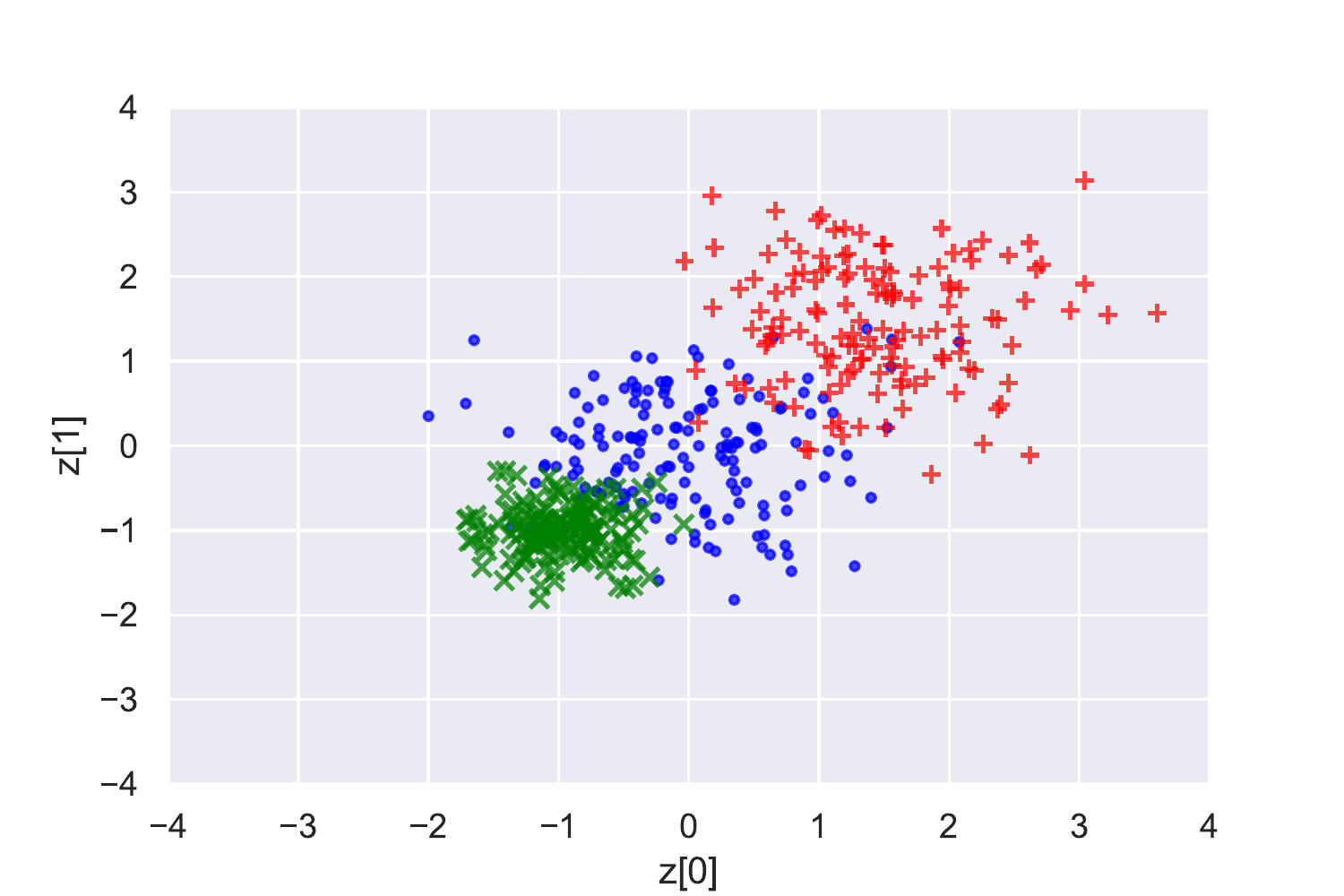}
        \caption{True embeddings}
    \end{subfigure}
    \\
    \begin{subfigure}[b]{0.45\textwidth}
        \centering
        \includegraphics[width=1\textwidth]{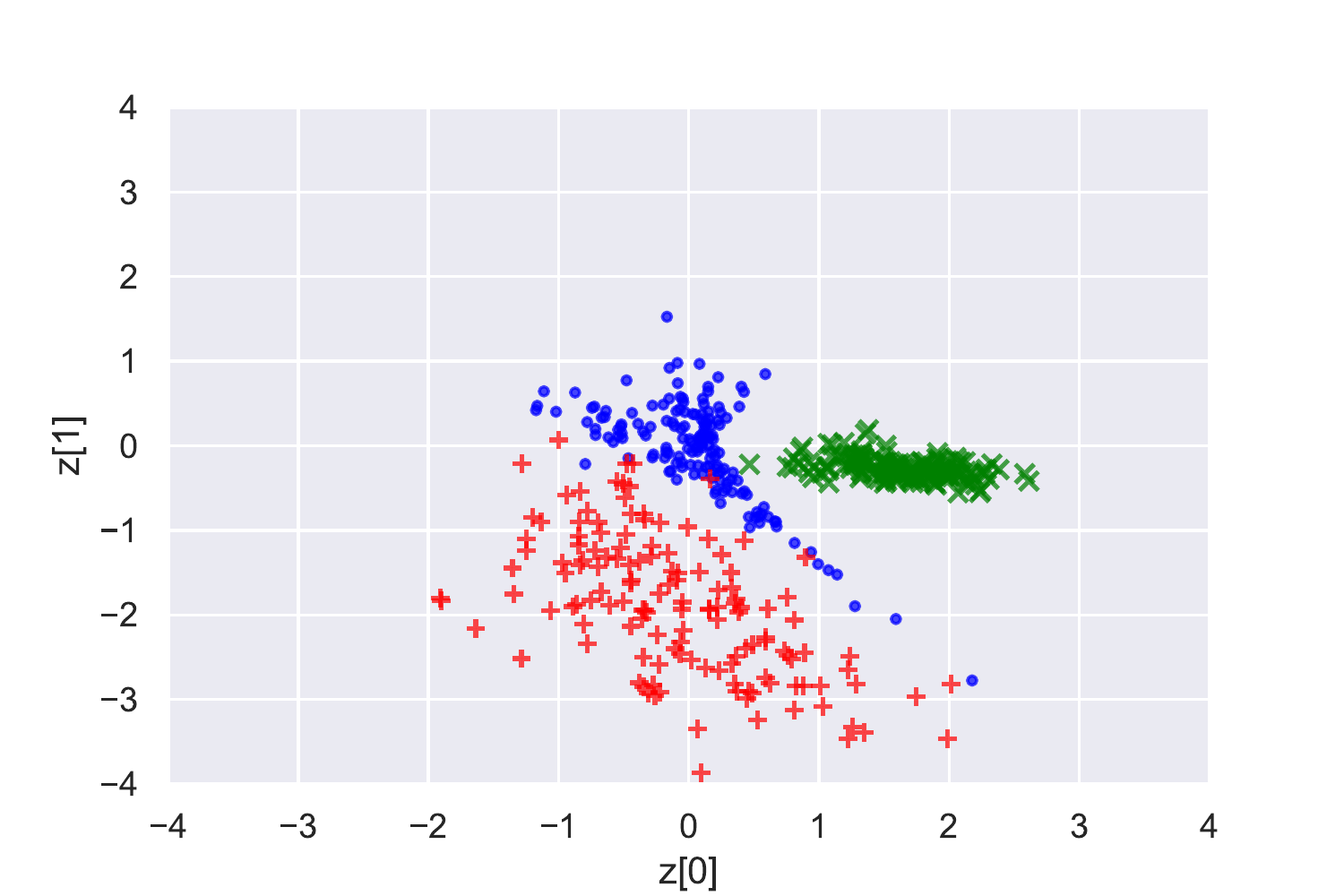}
        \caption{Proposed model embeddings}
    \end{subfigure}
    \caption{2D Latent space visualization of 100D count vectors}
    \label{fig:embed}
\end{figure}

\section{Experiments}

In this section, we perform numerical experiments to illustrate the proposed model. We start with simulation studies, then conclude with experiments on a bacterial microbiome dataset. 

\subsection{Simulations}

We generate synthetic datasets i) to explain the model selection strategy, ii) to demonstrate the accuracy of the latent embeddings, and iii) to show the ability to capture the covariance structure from observed data. 

\begin{figure*}[h]
    \centering
    \hspace*{-5em}
    \begin{subfigure}[b]{0.35\textwidth}
        \centering
        \includegraphics[width=1\textwidth]{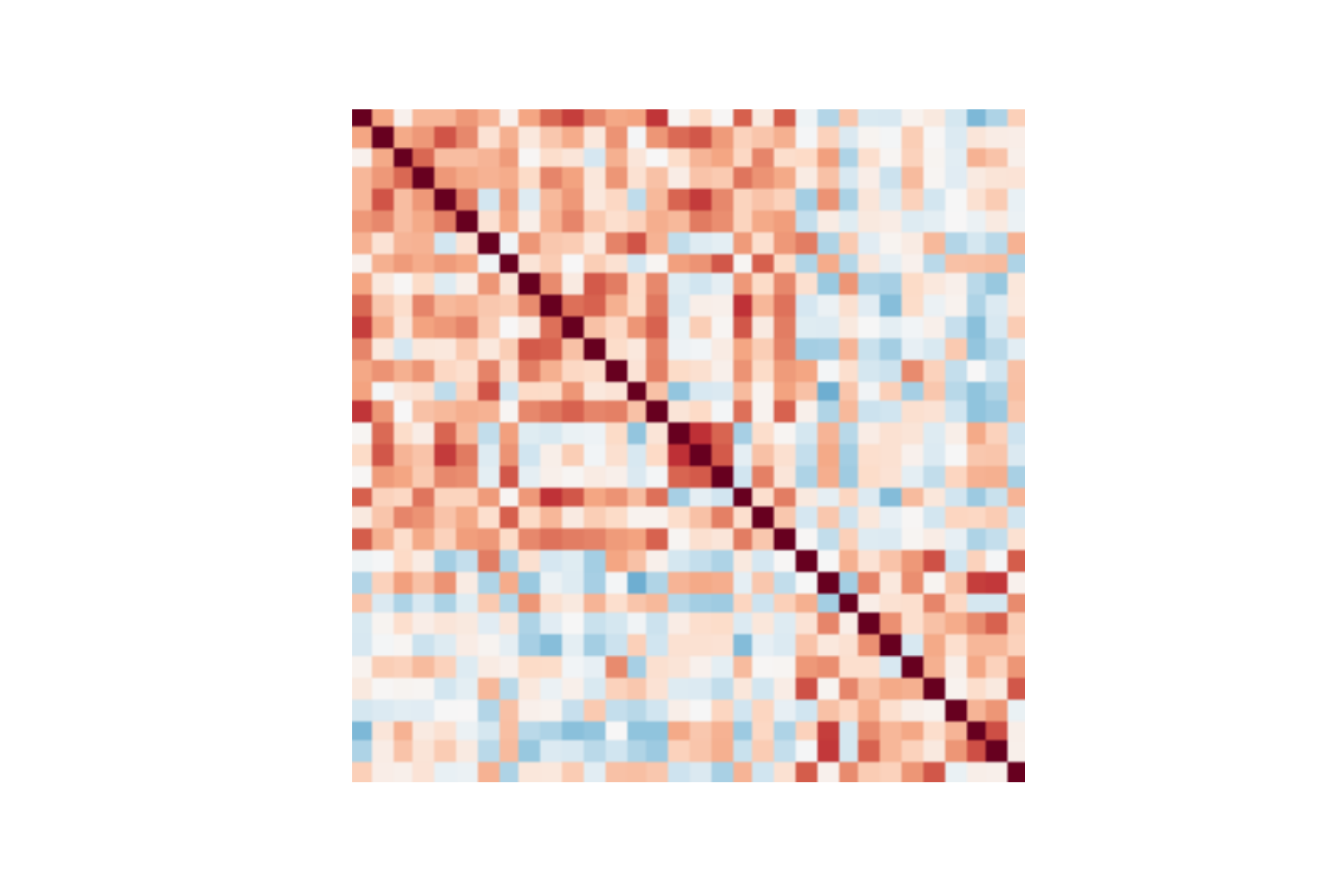}
        \caption{True Covariance}
        %  \label{fig:three sin x}
    \end{subfigure}\hspace*{-5em}
    \begin{subfigure}[b]{0.35\textwidth}
        \centering
        \includegraphics[width=1\textwidth]{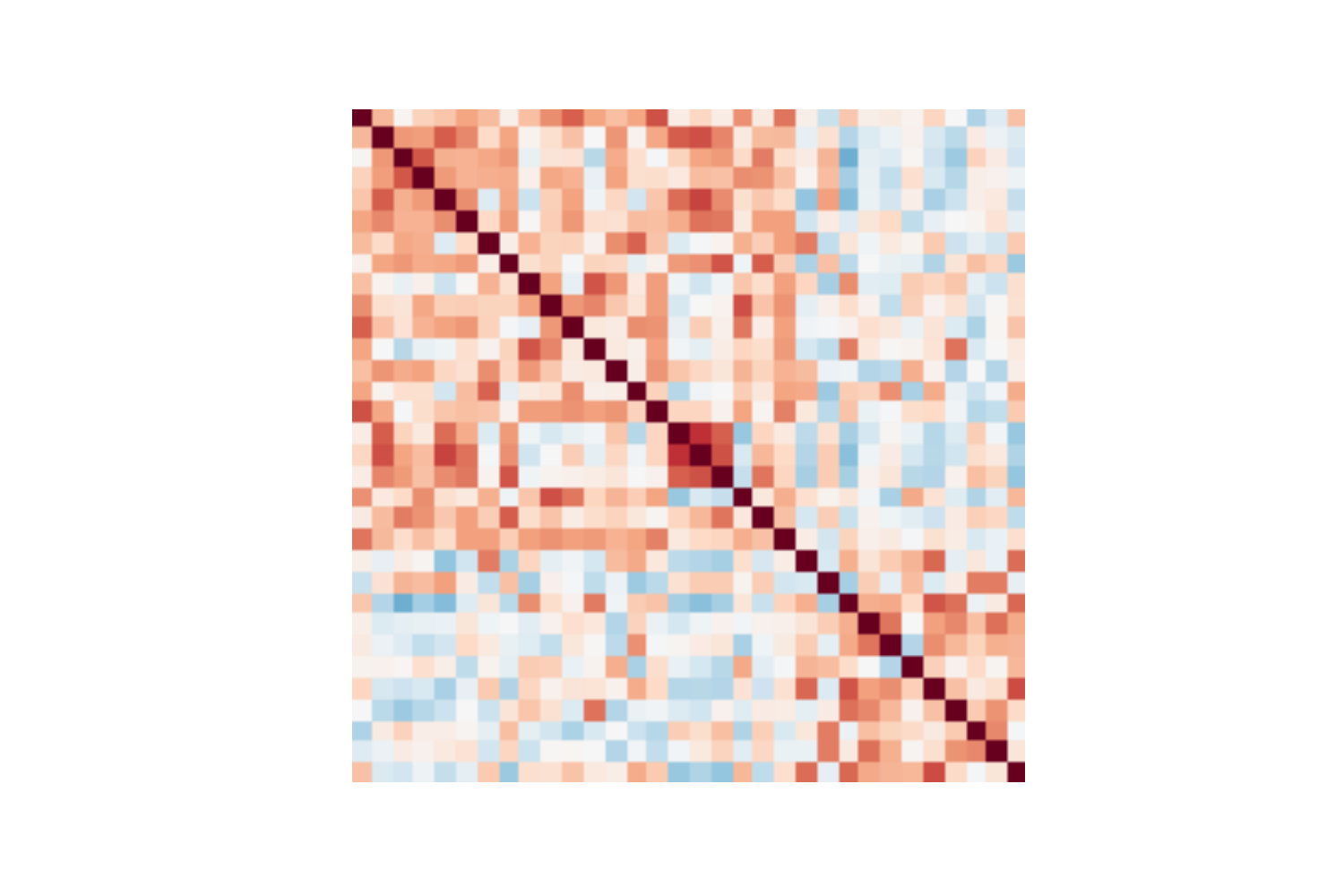}
        \caption{Proposed Model Covariance}
        %  \label{fig:three sin x}
    \end{subfigure}\hspace*{-5em}
    \begin{subfigure}[b]{0.35\textwidth}
        \centering
        \includegraphics[width=1\textwidth]{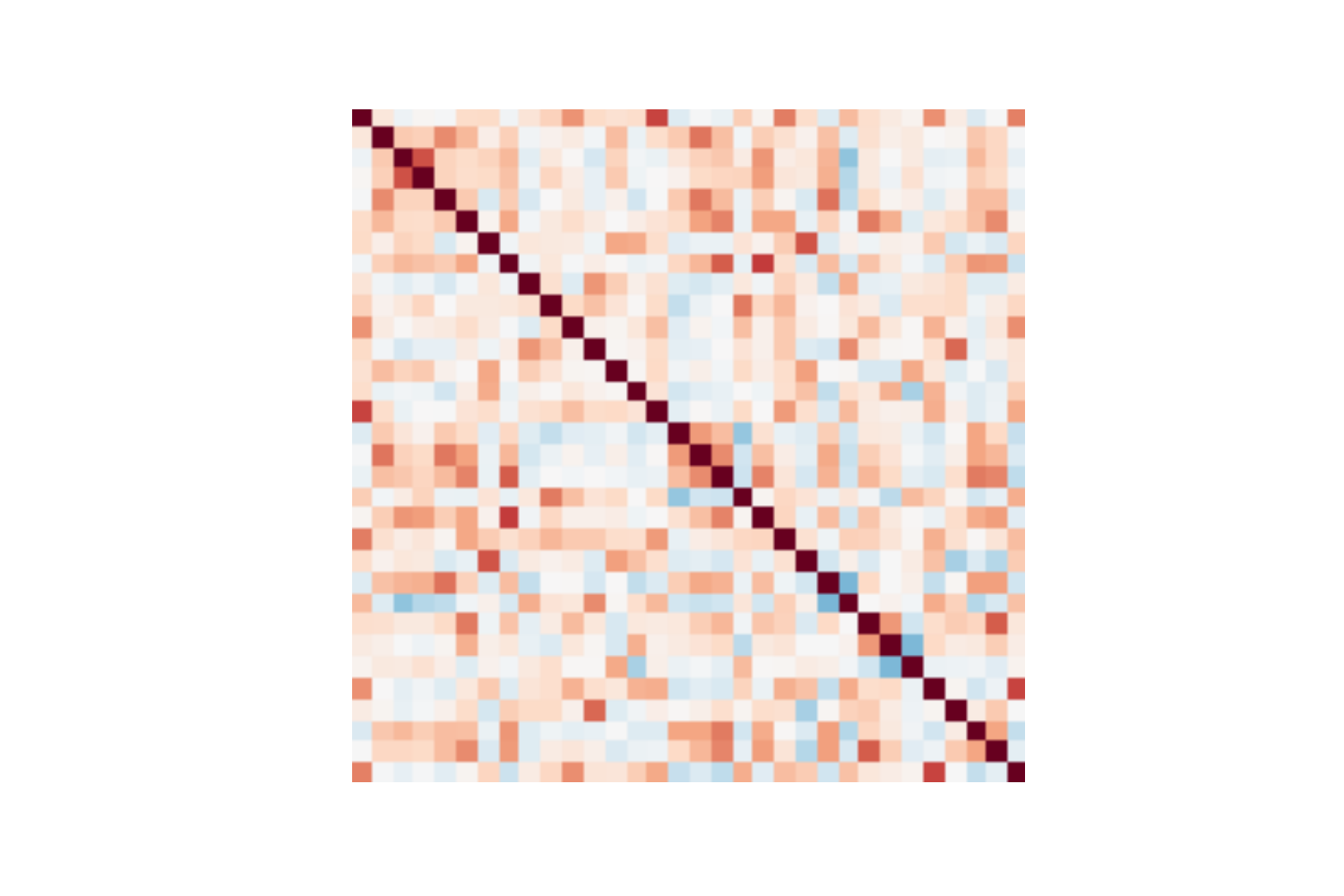}
        \caption{Graphical Lasso Covariance}
        %  \label{fig:three sin x}
    \end{subfigure}\hspace*{-5em}
    \\
    \hspace*{-5em}
    \begin{subfigure}[b]{0.35\textwidth}
        \centering
        \includegraphics[width=1\textwidth]{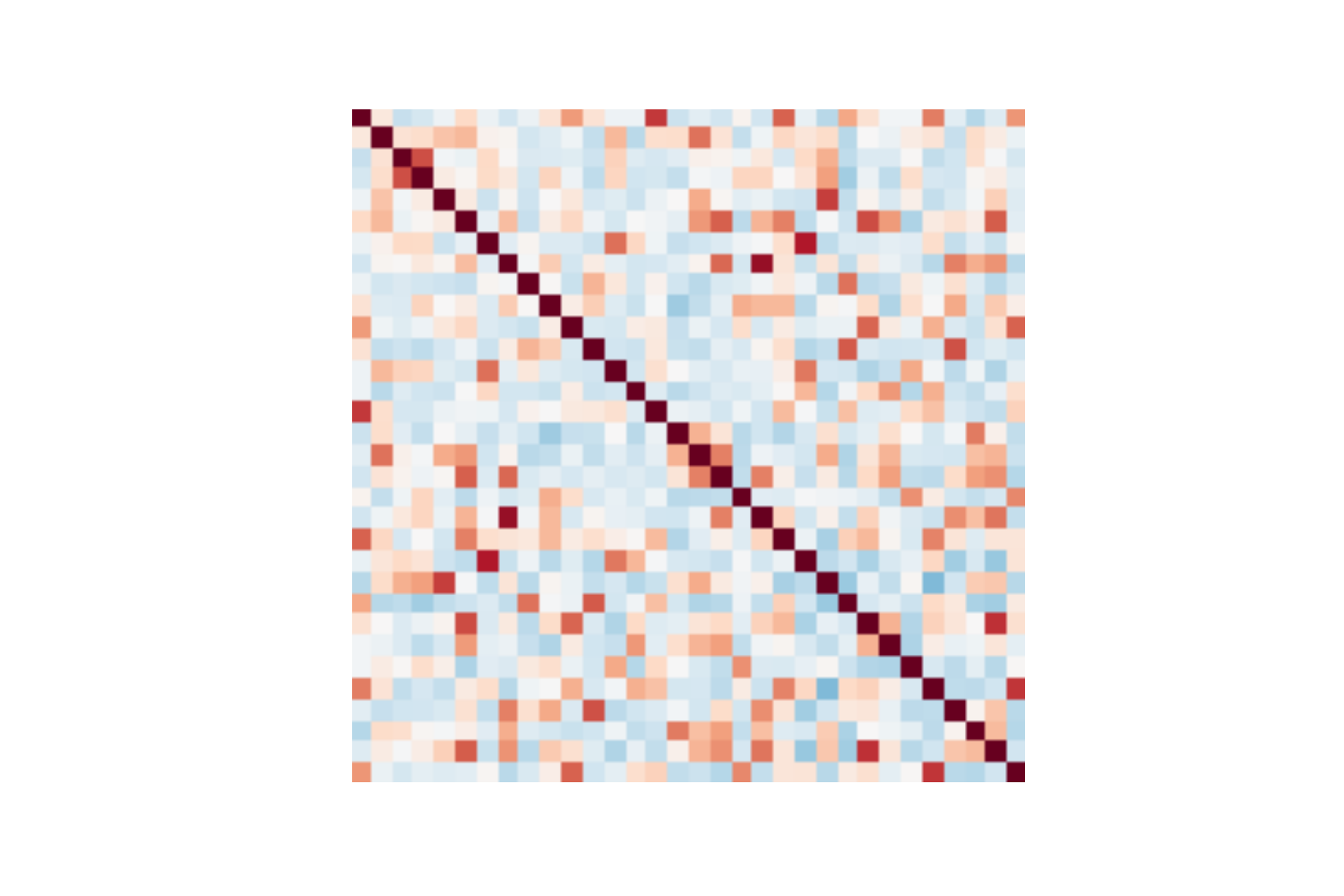}
        \caption{Emprical Covariance}
        %  \label{fig:three sin x}
    \end{subfigure}\hspace*{-5em}
    \begin{subfigure}[b]{0.35\textwidth}
        \centering
        \includegraphics[width=1\textwidth]{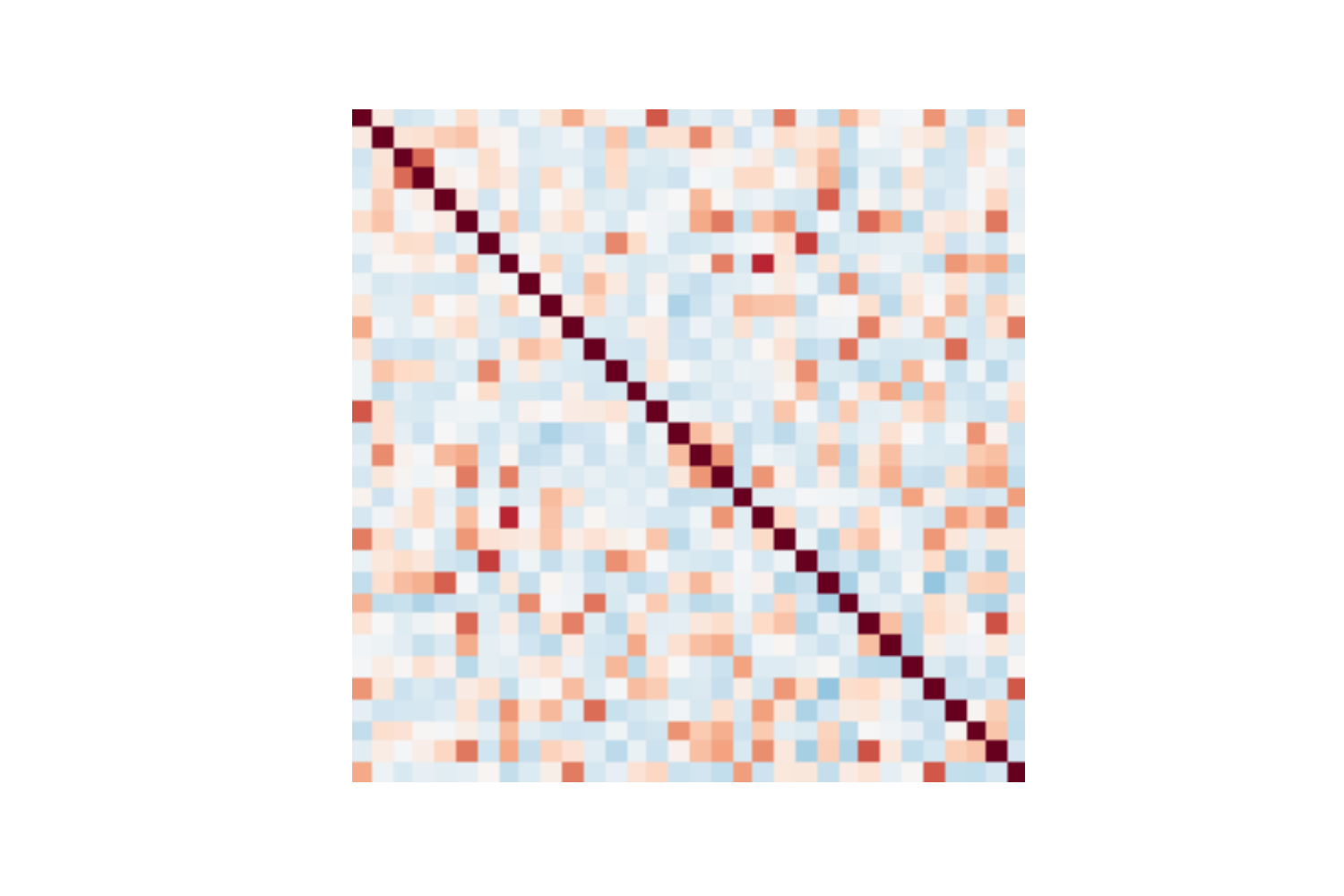}
        \caption{Ledoit-Wolf Covariance}
        %  \label{fig:three sin x}
    \end{subfigure}\hspace*{-5em}
    \begin{subfigure}[b]{0.35\textwidth}
        \centering
        \includegraphics[width=1\textwidth]{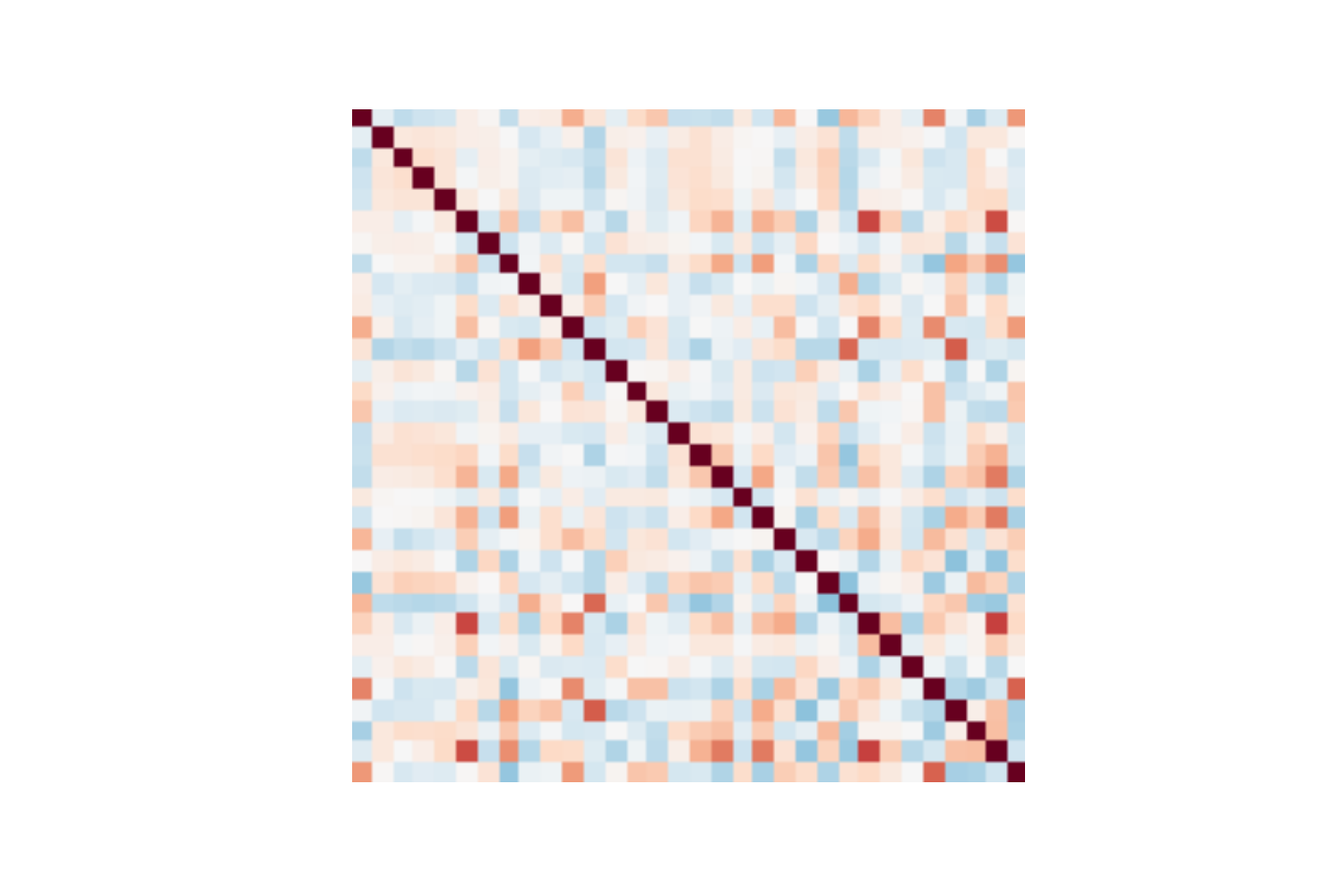}
        \caption{FA Covariance}
        %  \label{fig:three sin x}
    \end{subfigure}\hspace*{-5em}
    \caption{
    %\aoh{Mehmet: Only covariance matrices are mentioned in the subcaptions. So why is the word "Precision metrices" used in the caption?} 
    Estimated covariance matrices produced by the considered algorithms for a three species community with simulated transcript data. For more details on the model see Section \ref{sec:sim_single_com}. The proposed model provides a much more accurate estimated covariance than than do the other methods.}
    \label{fig:sim_multi}
\end{figure*}

\subsubsection{Model Selection} \label{sec:model_selection}
The proposed algorithm estimates the covariance matrix with a low-rank decomposition. The rank of the matrix is equal to the number of components $d_z$ in the latent space. which is a model hyper-parameter to be determined. We use the Bayesian Information Criterion (BIC) to estimate this parameter. The BIC arises from the Laplace approximation to the model posterior $p(M | D_k)$ \cite{konishi2008information}, where $M$ is the complete model including the latent dimension $d_z$. This results in a Bayesian estimate of $d_z$: $d_z = \text{argmax}_{d_z} ( \log p(D_{k}) - \text{BIC}/2)$, where BIC is a function of the total number of unknown parameters, which penalizes the log likelihood with a model complexity penalty term. In the proposed model, we use ELBO lower bound to the the likelihood by following \cite{beal2006variational}. The unknown parameters of the model are  $\{\Theta_{kl}\}_{l=1}^L$, $\bm{\mu}_k$, and $\bm{\Sigma_k}$. Hence, the total number of parameters is $\text{dof} = K \times (d_{l} +  K)$, which is used in the BIC expression as $0.5 \times \text{dof} \times \log(I_k) $. To illustrate the BIC model selection for the proposed model, we simulate three datasets with true latent space dimensions 4, 8, and 12, respectively, and then train multiple models while varying the dimensions $d_z$ over $\{2,3,\dots,12\}$ as the search range. We repeat the experiment 10 times to report the performance. The panel on top of Fig. \ref{fig:rank} shows the average BIC values obtained after convergence of the variational EM algorithm. We see that maximum BIC is obtained on the vicinity of the true ranks for all the datasets. On the other hand, the panel on the bottom shows the RMSE values of the estimated covariance matrices. One can see that the lowest error is achieved by maximizing $d_z$, which is likely to result in over-fitting the model. This over-fitting is mitigated by using BIC estimate of $d_z$.

\subsubsection{Embedding Characteristics}
We generate a synthetic dataset with a 2 dimensional latent space having 3 different classes, i.e., experimental conditions, according to the model specification in Section \ref{sec:model}. The latent variables are sampled for each class from different Gaussian distributions. The associated means are predefined as $[0,0], [1.5,1.5], [-1,-1]$ and the variances of the isotropic covariances are selected as $0.5, 0.5, 0.1$, respectively. Three class conditional densities are generated with different affine transformation parameters. The observation space is 25 dimensional. The observations are sampled from the conditional multinomial distributions with soft-max link function as in Eq. \ref{eq:x_pr}. We generate 200 observations for each class with fixed total number of counts, which is 100, per observation, then stack all the observations. Fig. \ref{fig:embed}.a shows the true embeddings of the resulting dataset. We trained the proposed algorithm with the true latent space dimension. Fig. \ref{fig:embed}.b shows the embeddings of the model, which are obtained through the posterior distributions. Due to non-identifiability of the model, the latent variables can only be recovered up to a rotation. The distorted shape of the latent clusters in  Fig. \ref{fig:embed}.b is due to the use of the soft-max link function. If there is a large component in the affine transformed latent vector, the other components are washed out, hence such points would map to very close points in the observation space. Notwithstanding the differences between Fig. \ref{fig:embed}.a and Fig. \ref{fig:embed}.b, the model preserves the clustering structure accurately.  

\subsubsection{Influence of the Total Counts and Dimensions on Performance} \label{sec:counts}
The number of counts of the observed vector $\bm{x}_{kl,i}$ is an observation-specific parameter, which affects the accuracy of the proposed algorithm. Figure \ref{fig:counts} shows the effect of the number of counts $N_{kl,i}$ on the RMSE values of the covariance estimator under three different latent dimension settings. We sample the total counts of a simulated vector from the Poisson distribution with fixed mean.  %Hence, each observation will  have somewhat different total counts. 
We also fix the observation dimension to 128. RMSE is reported based on averaging 20 experiments. Figure \ref{fig:counts} shows that increasing the mean number of counts improves performance. In particular, we see that the total counts $N_{kl,i}$ and the mean error are inversely proportional. This is expected since the number of counts directly affects the posterior uncertainty (Eq. \ref{eq:post_cov}) and mean (Eq. \ref{eq:post_mean}). The contribution to the ELBO of the observations increases as the total count increases. Furthermore, for low number of counts, the covariance matrix becomes harder to predict due to higher vulnerability to over-fitting. On the other hand, Fig. \ref{fig:dims} demonstrates the opposite trend when the dimension of  the observation space dimension is increased. Here the total mean counts is fixed to 100. In higher dimensional datasets, the model struggles to estimate the covariance structure when the rank is low. However, this phenomenon diminishes when we observe more counts as can be seen in Fig. \ref{fig:counts}.

\subsubsection{Baseline Algorithms}
% Algorithms
Here we show performance comparisons of the proposed method relative to four baseline methods for estimating the underlying covariance and inverse covariance matrices. i) Empirical covariance, which is the sample covariance. ii) The Ledoit-Wolf estimator \cite{ledoit2004well}, which uses shrinkage regularization to perform MAP estimation for the covariance matrix by assigning an inverse Wishart prior on the covariance matrix. iii) Gaussian Copula GraphicalLasso \cite{liu2009nonparanormal}, which penalizes the covariance matrix in a different way, particularly, with L1-norm constraints on the precision matrix after transforming the data by using Gaussian copulas. Regularization forces the entries of the precision matrix to be sparse. iv) Factor Analysis \cite{murphy2012machine} uses another form of regularization of the covariance matrix by imposing low-rank structure. Each of these baseline methods is expected to perform best when the data, or its transformed version, is normally distributed. 
%Note that the baseline algorithms assumes normality of the observations. 
For the Ledoit-Wolf, GraphicalLasso, and FA, as is customary, we first normalize the data by subtracting the mean and dividing by the variance, before running these methods. On the other hand, as the non-Gaussian counting nature of the data is explicitly modeled in our proposed model, our algorithm is run on the raw observations. For model selection in both FA and proposed model, we use the exact rank of the simulated dataset. The regularization coefficient of Gaussian Copula GraphicalLasso algorithm is estimated by using 5-fold cross-validation. For the Ledoit-Wolf algorithm, we used the expression for the shrinkage coefficient given in \cite{ledoit2004well}. We use {\tt scikit-learn} implementations of the baseline methods.  

\subsubsection{Simulating Model Communities} \label{sec:sim_single_com}
Next, we generate a synthetic dataset which contains the transcript abundance data (an estimate of gene expression) of two different species existing in the same community, hence $L=2$.  The latent variables $\bm{z}_{i}$ with dimension $d_z = 5$ are generated for each measurement site by sampling from $\bm{z}_{i} \sim \mathcal{N} (\bm{0}_{d_z} , \bm{I}_{d_z})$, where $i$ indexes the replicate for $i=1:I$. These latent variables have elements that correspond to the hidden factors generating the data, such as environmental variables, mediator species effects, and direct associations. We transform the latent variables to the probabilities in the observation space, whose dimensions (abundance of transcripts) are chosen as $d_1 = 20, d_2 = 10$ by using affine and subsequently soft-max transformations as described in Section \ref{sec:model}. The parameters $\Theta_l \in \mathbb{R}^{d_l\times d_z}$, are chosen randomly by sampling from a zero mean multi-variate normal distribution. Then, we sample the observed data $\bm{x}_{l,i}$ from the multinomial distribution. The total counts $N_{l,i}$ of a sample is chosen randomly by sampling from a Poisson distribution with rate parameter $1000$. We simulate a total of $I = 200$ replicates for each dataset. The true covariance matrix is then given as $\tilde{\Theta}_{l}\tilde{\Theta}_{l}^T$, where $\tilde{\Theta}_{l} =  [ \Theta_{1}, \Theta_{2}]$.

\subsubsection{Correlation Results}
% Results
%\aoh{I don't think we need both figures and suggest that you eliminate the first one - the figures tell the same story and one is sufficient. I assume that the second figure is for 3 species? You should say so and mention in the caption how the species are related to the blocks in the true covariance.}
Fig. \ref{fig:sim_multi} show the estimated covariance matrices of the baseline algorithms, the proposed algorithm, alongside with the ground truth matrix, when simulated dataset is realized. The proposed model can recover the covariance structure accurately. The relatively poorer accuracy of the other methods can be attributed  to several factors. First, these models do not exploit the counting nature of the data. The second reason is that the covariance matrix is simulated with low-rank structure, which is not taken into account by the Gaussian Copula Graphical-Lasso, Ledoit-Wolf, or standard sample covariance estimation methods. %estimate full-rank covariance matrices.
As the data was simulated from the proposed model, the proposed algorithm naturally performs better. 
%better models the data by explicitly assuming a multinomial distribution while imposing low-rank constraint on the covariance matrix. 
Note also that, for multiple species, the proposed model can discover both inter-species and intra-species correlations. Table \ref{tab:sim} shows the resulting RMSE values between the estimated and the ground truth covariance matrices for the aforementioned simulation setting. The proposed model achieves lower error in overall. This is expected since the model uses an ELBO approximation to the true marginal likelihood function. %Note that, in the multiple communities with covariates case, it is clear that the algorithms other than the proposed one performs worse in estimating covariance matrix when there are control variables (compare 'Multi' and 'Covariate' columns for Covariance estimation). The proposed algorithm can maintain its performance by effectively modeling the conditional joint density given the control variables. 

\begin{table}[]
    \centering
    \small
    \begin{tabular}{llcc}
        \hline
        &\textbf{Algorithm} & \textbf{Covariance} & \textbf{Precision}\\ 
        \hline
        &Empirical            & .231 $\pm$ .013 & .510 $\pm$ .021 \\ 
        &Ledoit-Wolf          & .225 $\pm$ .011 & .150 $\pm$ .004\\ 
        Cov &FA               & .237 $\pm$ .014 & .086 $\pm$ .003\\ 
        &GLasso               & .173 $\pm$ .013 & .073 $\pm$ .002\\ 
        &Proposed             & .096 $\pm$ .022 & .003 $\pm$ .001\\ 
        \hline
    \end{tabular}
    \caption{Mean and standard deviation of RMSE between the estimated covariance matrices and ground truth over 10 different realizations of the simulated abundance dataset. }
    \label{tab:sim}
\end{table}

%\subsection{Real-world datasets}
\subsection{Bacterial Community Experiment} \label{sec:mic}

In this section, we demonstrate a real world use-case of the proposed model: transcript analysis of a bacterial model community called THOR \cite{hurley2022}. 

%\subsubsection{Analysing ITS/16S Microbiome Communities} \label{sec:mic}

Microbial model communities are useful to understand principles that govern community behaviours \cite{BENGTSSONPALME20203987, https://doi.org/10.1111/1462-2920.12343, doi:10.1128/mSystems.00175-19, doi:10.1128/mSystems.00161-17}. The Hitchhikers Of the Rhizosphere (THOR) is a model community consisting of three microbial species, {\it Bacillus cereus, Flavobacterium johnsoniae}, and {\it P. koreensis} that co-isolate from field-grown soybean roots. The organisms in THOR represent three dominant rhizosphere taxa (at the phylum level), and are common in soil and the mammalian gut.  
%These are the three dominant Rhizosphere taxa, which frequently exist in soil, mammalian gut, and soybean environments. 
{\it B. cereus} is a Firmicute that carries {\it F. johnsoniae}, a member of the Bacteriodetes, and  {\it P. koreensis}, a member of the Proteobacteria, as hitchhikers \cite{doi:10.1128/mBio.02846-18}. Due to their abundance in several environments, their may demonstrated interactions in the lab and field, and their genetic tractability, these species make a useful model community with relevance to the natural world. The model community provides a simple system in which to study and model community level interactions, which are poorly understood. Developing governing principles of community behavior may lead to strategies to manipulate microbiomes for human or environmental health. 
%dominating existence of these species in several conditions, it is of significant importance for microbiologists to infer community level interactions to be able to understand and modify emergent behaviours, and consequently improve human health, environmental sustainability, agricultural productivity, etc \cite{SHETH2016189}.

\begin{figure*}
    \centering
    \begin{subfigure}{0.33\textwidth}
        \centering
        \includegraphics[width=1\textwidth]{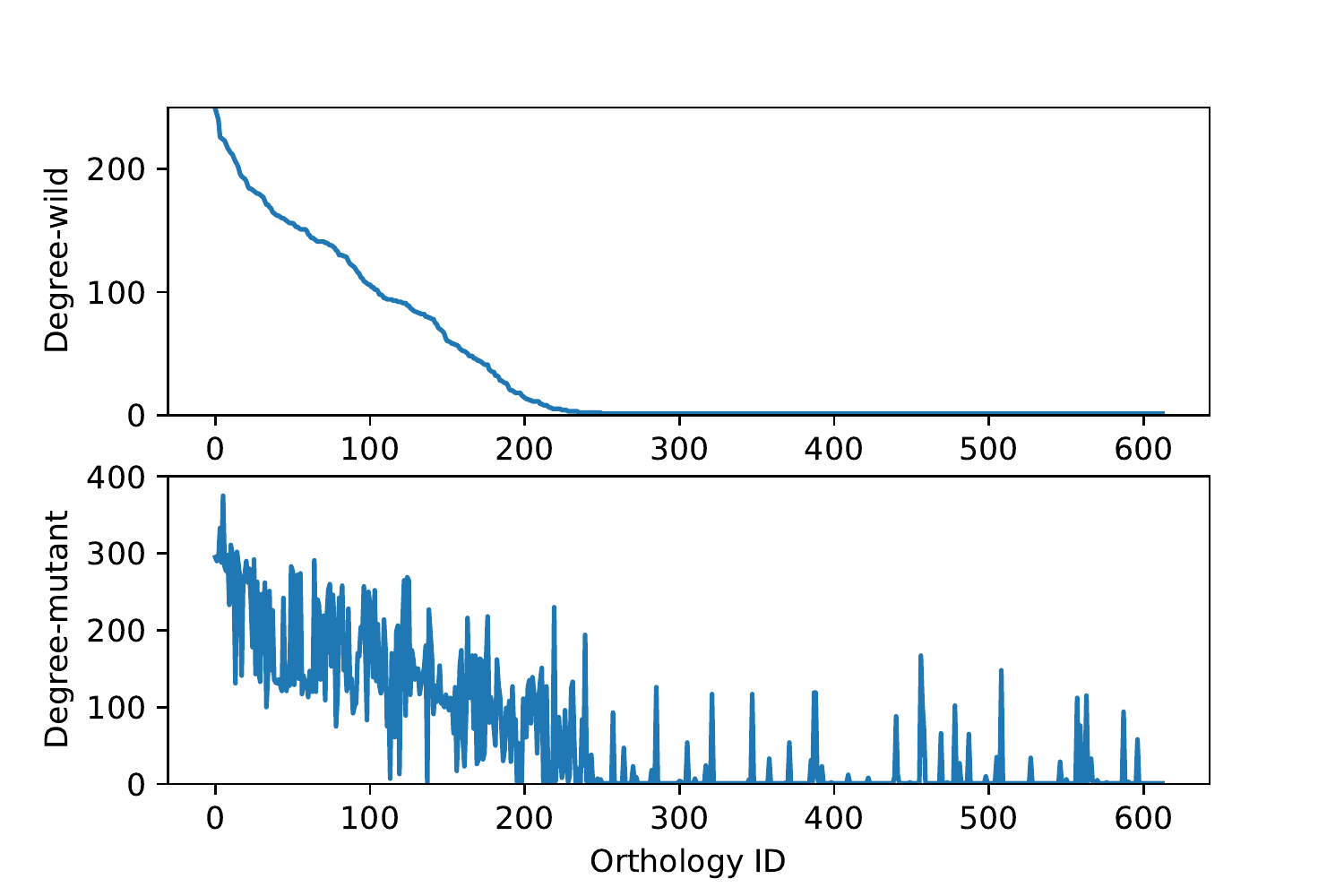}  
        \caption{B. cereus}
    \end{subfigure}\hfill
    \begin{subfigure}{0.33\textwidth}
        \centering
        \includegraphics[width=1\textwidth]{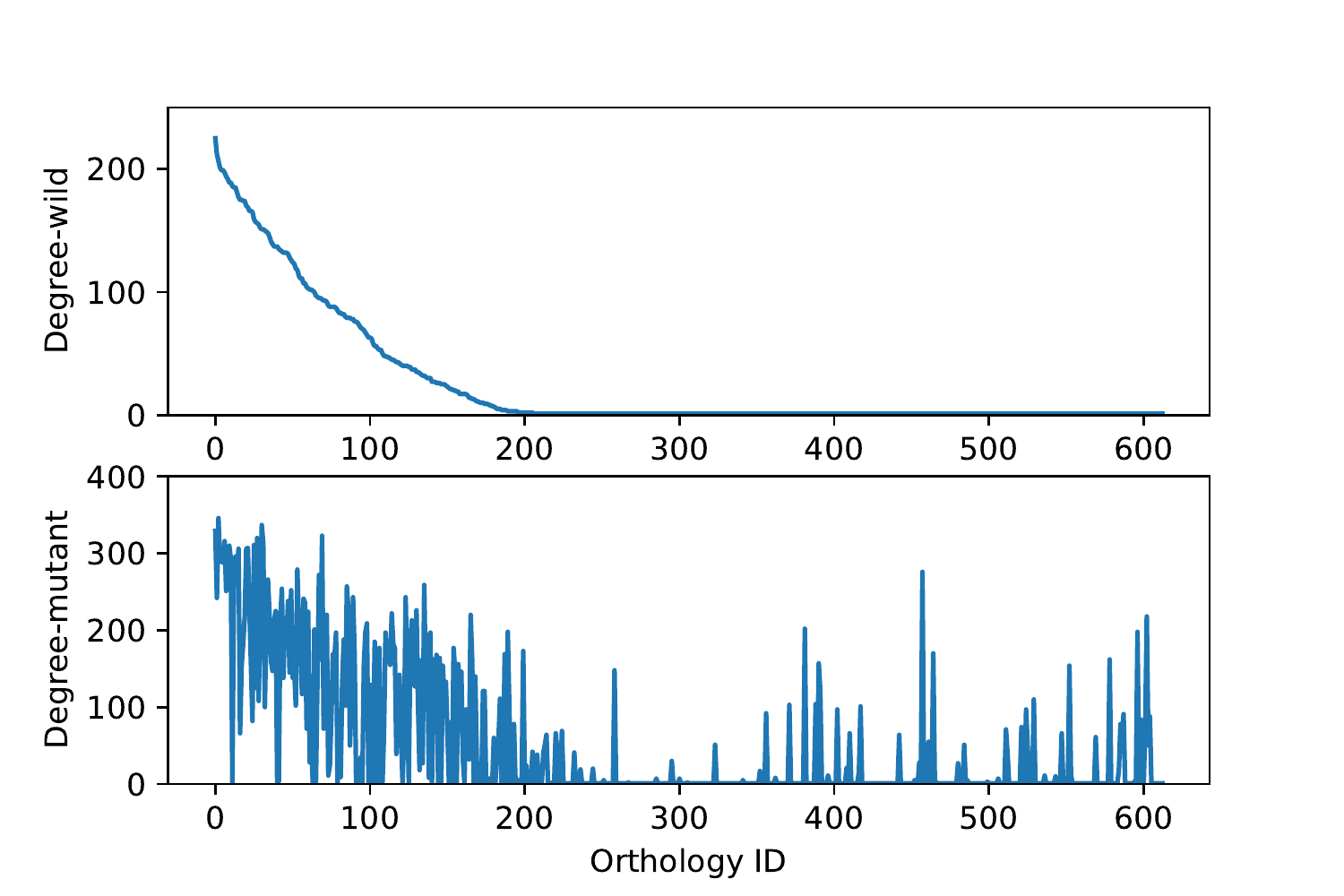} 
        \caption{F. johnsoniae}
    \end{subfigure}
    \begin{subfigure}{0.33\textwidth}
        \centering
        \includegraphics[width=1\textwidth]{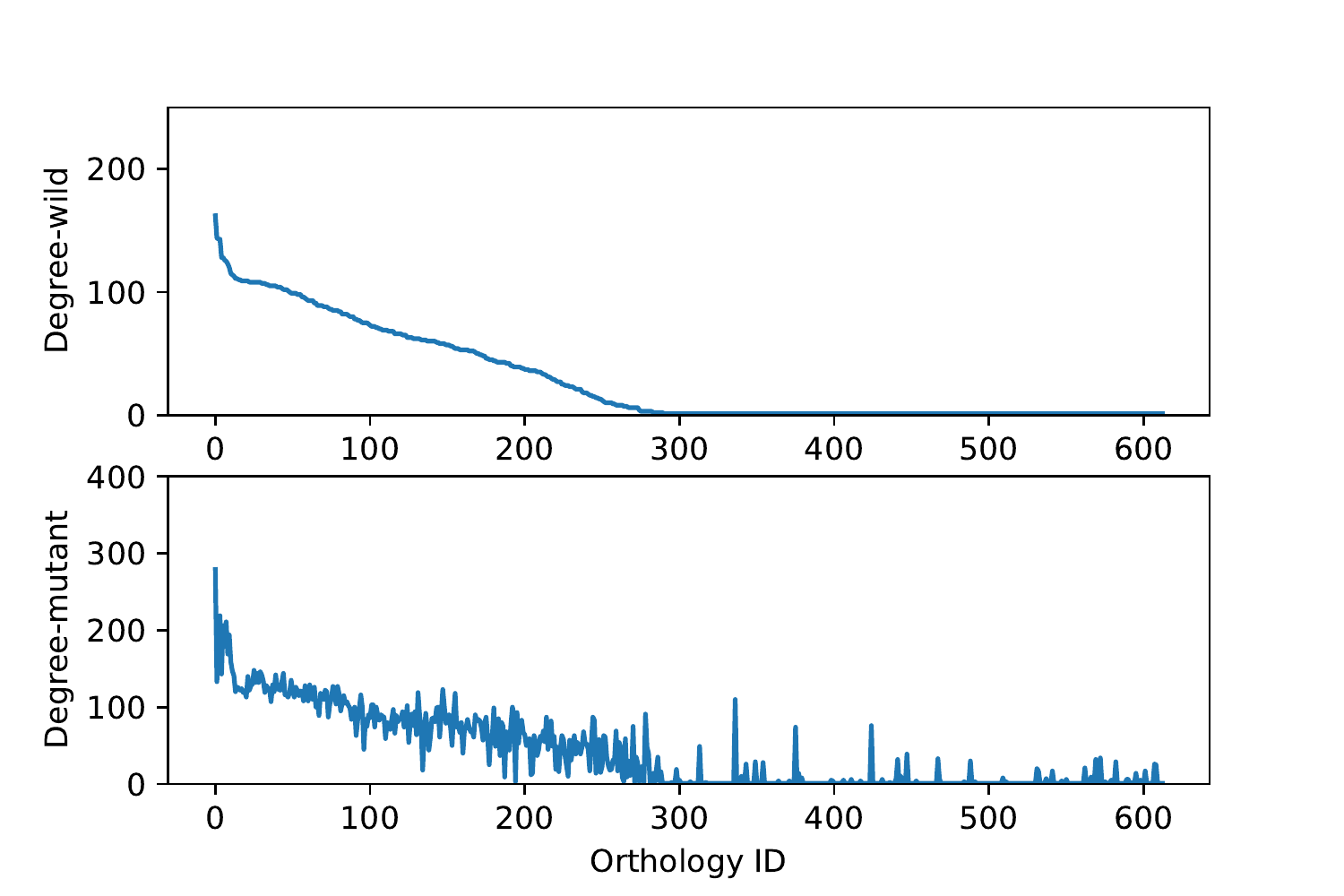} 
        \caption{P. koreensis}
    \end{subfigure}
    \caption{Effect of koreenciene removal on the the centrality (vertex degree) of vertices in the transcriptional orthology correlation networks inferred from our model for the experimental THOR dataset.  For each species, the ortholog IDs are sorted in decreasing order of the wildtype vertex degree.  The upper row shows plots of the degree of each vertex (transcriptional orthology ID), in descending order of magnitude, for the wildtype condition. The bottom row shows  corresponding plots of the vertex degree when the koreenceine pathway is removed (mutant condition), under the same ordering of vertices as in the top row. {\it P. koreensis} preserves its network connectivity better than the other two species. The network connectivity of  {\it F. johnsoniae} is the most affected by koreenciene removal. }
    \label{fig:degree_change}
\end{figure*}

The dataset is collected under two conditions associated with the treatments applied to {\it P. koreensis}. In the first condition the THOR community contains the wild type {\it P. koreensis} strain and in the second condition the wildtype is replaced with a mutant of {\it P. koreensis} that does not produce koreenceine antibiotics. Production of koreenceines is an important factor in community interactions because they inhibit growth of {\it F. johnsoniae} \cite{doi:10.1128/AEM.03058-18} and {\it B. cereus} protects {\it F. johnsoniae} by modulating koreenceine levels. 
%
%{\it P. koreensis} inhibits the growth of {\it F. johnsoniae} \cite{doi:10.1128/AEM.03058-18}. Indeed, the produced Koreenceine metabolites are the main inhibitor. However, B. cereus protects F. johnsoniae when it exist in the model community by modulating the levels of koreenceine metabolites. 
%
By using our proposed model, in particular the associated estimated joint probability density of the data, we will be able to reveal effects of the treatment. Since the joint probability density model is parameterized by the mean and covariance of a multivariate Gaussian latent variable (See Section \ref{subsec: correlation analysis}), the mean and covariance parameters play the principal role in our metatranscriptomic analysis. For brevity we focus our discussion on the inferred covariance parameters here (See Supplementary for discussion of the mean parameters inferred by the model).  %\aoh{I don't see where 5the mean parameters are discussed here.}\\\\\\\\\\\\\\\\

The microbial community dataset consists of a total of 17244 gene transcripts associated with three species. There were respectively 38 and 36 replicates for the community with wildtype and mutant strains of {\it P. koreensis}. 343 transcripts were removed from the analysis as they had zero counts over all experimental replicates.  After removing these transcripts, {\it B. cereus, F. johnsoniae,} and {\it P. koreensis} express 5903, 5146, 5852 transcripts, respectively. 
We reduced the dimension of the feature space using orthological groupings of gene transcripts into metabolic  pathways\footnote{The transcriptional orthology mappings of the THOR gene transcripts to metabolic pathways were obtained using Kegg \url{https://www.genome.jp/kegg/}. See supplementary for an example.}.   
%form  data structure of features 
% from Bc and Pk assemblies publicly available on DOE's JGI genome portal \url{https://genome.jgi.doe.gov/portal/}.}.
Specifically, after pathway mapping each feature corresponds to a transcritpional orthology ID, and the associated data is the summation of the counts of the transcripts tagged with that ID. We aggregated all the transcripts that were not mapped to any Kegg ortholog into a single non-assigned orthology ID, denoted KXXXXX, and we only considered those ortholog IDs that are present in all 3 species. This filtering resulted in a set of 613 ortholog IDs, which corresponds to the dimension of the feature space used in our model.
%. The dataset includes two experimental conditions, wild-type and mutant based on the treatment applied to P. koreensis. The total numbers of the replicates for wild-type and mutant conditions are 38 and 34, respectively.

The rank of the proposed model was determined by successively fitting the model to  latent spaces of dimensions ranging between 5 and 50 with increments of 5.
%\aoh{Mehmet: Is this correct? If not please insert the right numbers.}
Then, the optimal model rank was determined as the latent dimension that yields the highest value of the BIC as described in Section \ref{sec:model_selection}. The optimal model rank was found to be 40. The parameters (mean and covariance) of the models were subsequently refitted with the optimal dimension. The probability distribution of the data is computed under the wildtype and mutant conditions, whose explicit form is given in Eq. \ref{eq:incuded_dist} as a marginalization over the latent variables.

\noindent{\bf Network centrality changes}: We evaluate the effect of removal of koreenciene (mutant) on the centrality of the inferred $613 \times 613$ correlation network of metabolic pathways. Here the centrality of a vertex of the network is measured by vertex degree, i.e., the number of edges connecting the vertex.  To ensure that the networks contain only the most biologically significant edges in the networks, we applied a very high correlation threshold ($0.95$) to the respective inferred wild-type and mutant correlation matrices produced by fitting our proposed graphical model to the data. Using such a high threshold is in line with established RNA-Seq network inference practices \cite{liesecke2018ranking}. 
% On p. "For these four normalization methods, thresholds used to get the best co-expressed gene lists were >0.9." 
% after thresholding,  %marginal distrmibutions between the transcriptional orthologies from which we can extract correlation matrices for wildtype and mutant (removal of koreenciene) conditions by normalizing by their diagonal entries. 
%According to the theory in \cite{}, this threshold is substantially larger than the phase transition thresholds $0.63$ and $0.6$, respectively, for $613 \times 613$ wildtype and mutant sample correlation matrices estimated from and $36$ and $38$ samples.  
%We are interested in the effect of the treatment on the network. To this end, we compute degrees of each orthology by applying a threshold. Particularly, only the correlations under -0.95 and above 0.95 are considered when computing the degrees. 
Figure \ref{fig:degree_change} illustrates the effect of removal of koreenceine on the degrees of the nodes (transcriptional orthology IDs) in these networks. Comparison of the upper panels with the lower panels of the figure indicates that the vertex degree distribution of {\it F. johnsoniae} is most affected, followed by {\it B. cereus}, with {\it P. koreensis} the least affected. %, although the mutation is applied to this species.
This relative ordering of sensitivity of the three species to koreenceine removal shown for vertex degree in Fig. \ref{fig:degree_change} mirrors the relative ordering of sensitivity shown for the mean changes (See Fig. 2
and associated discussion
%\ref{fig:mean_hist} 
in the Supplementary).   
%\aoh{This last sentence can be removed if we remove the mean analysis as suggested.}
%This is unsurprising since we expect the removal of koreenceine metabolites to effect this species. 

\begin{figure*}
    \centering
    \begin{subfigure}{0.33\textwidth}
        \centering
        \includegraphics[width=1\textwidth]{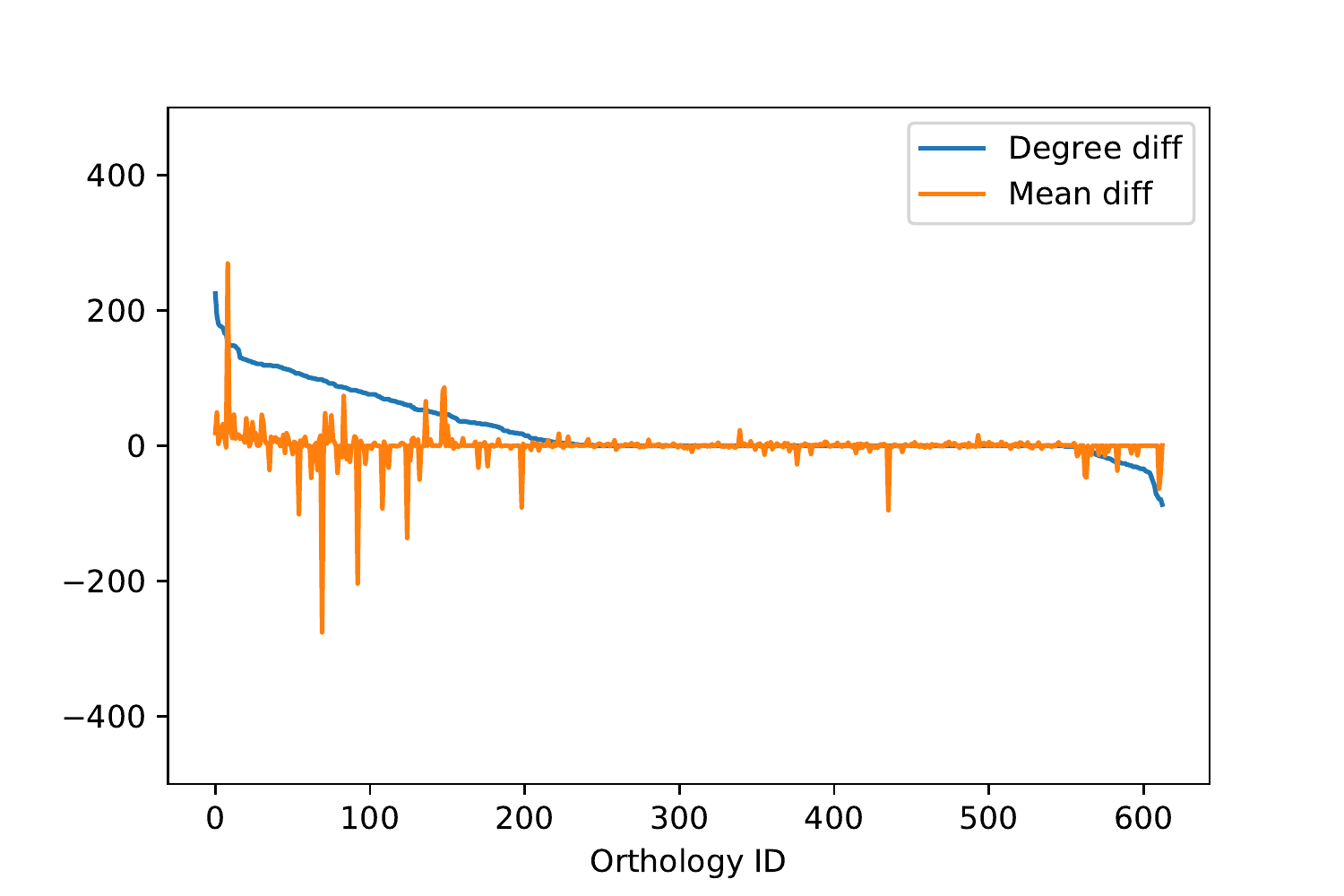}  
        \caption{B. cereus}
    \end{subfigure}\hfill
    \begin{subfigure}{0.33\textwidth}
        \centering
        \includegraphics[width=1\textwidth]{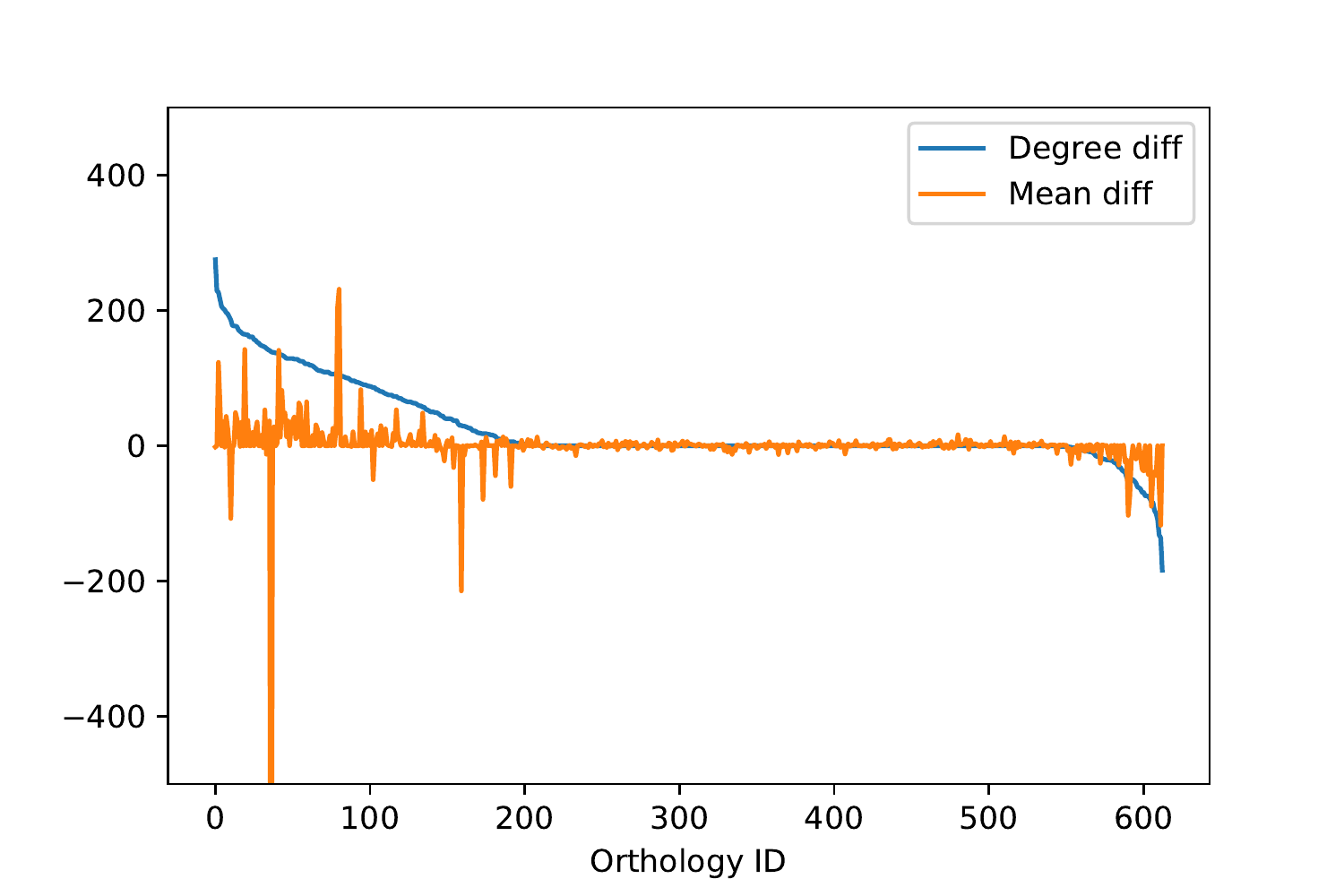} 
        \caption{F. johnsoniae}
    \end{subfigure}
    \begin{subfigure}{0.33\textwidth}
        \centering
        \includegraphics[width=1\textwidth]{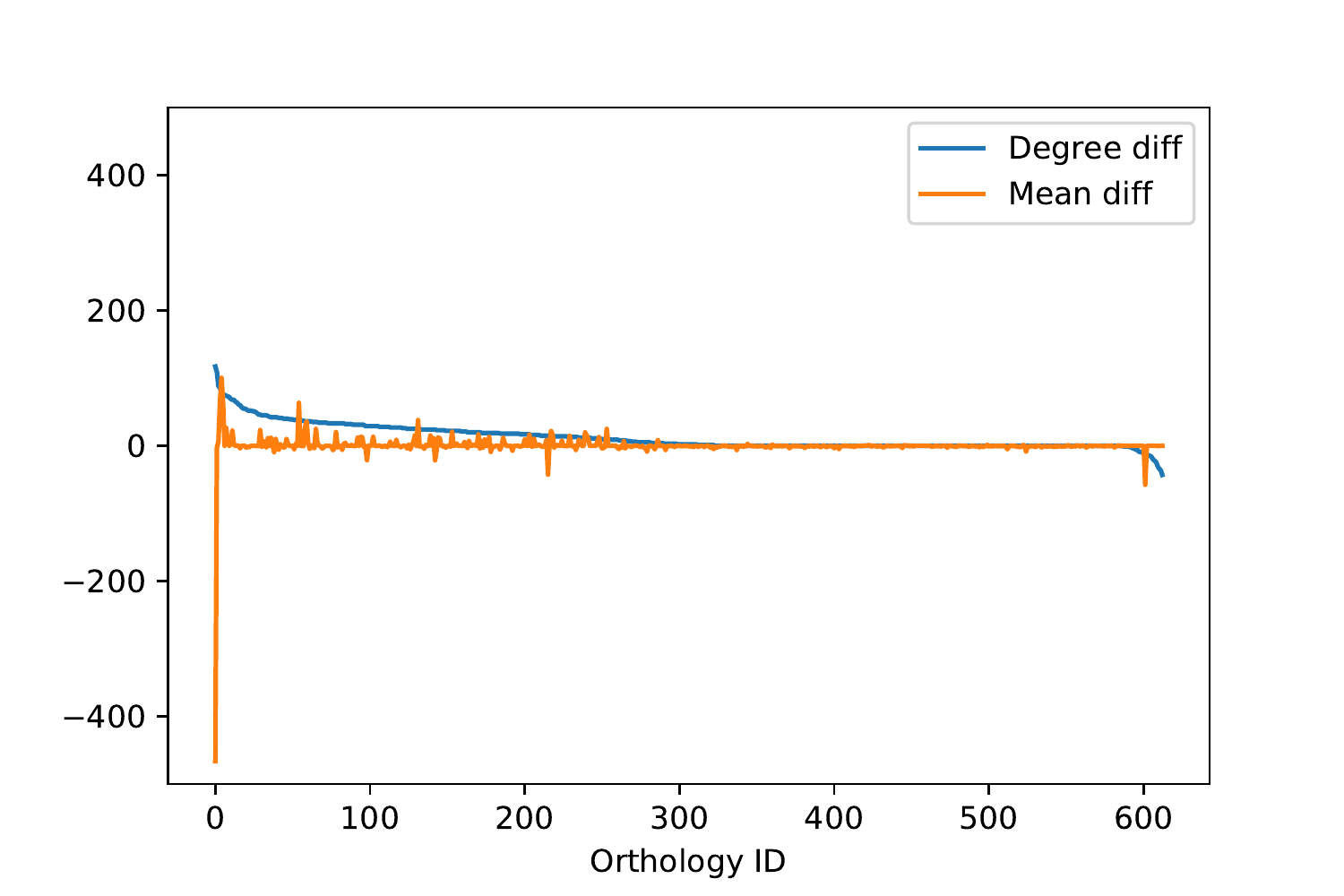} 
        \caption{P. koreensis}
    \end{subfigure}
    \caption{Effect of koreenceine removal on vertex centrality and vertex mean counts for the transcriptional orthology correlation network. For each species, the ortholog IDs are sorted in decreasing order of the vertex degree difference between mutant and wildtype. It is notable that, with few exceptions,  all  orthology IDs with significant changes in vertex mean also have changes vertex degree, but not conversely.  Furthermore, the asymmetry of the blue curve suggests that the removal of koreenceine is associated with an increase in network connectivity (many more vertices whose degrees increase than decrease), especially in {\it F. johnsoniae}.     
    %The differences are compatible, however reveal complementary information. P. koreensis preserves its mean and covariance structure best when the treatment is applied. F. johnsoniae is the most sensitive species under the treatment. Both mean and covariance structure change substantially.
    }
    \label{fig:degree_mean_change}
\end{figure*}

% Compatibility between mean and degree
%and mean changes together to visualize the different effects of the treatment on the mean and covariance structures. 
Fig. \ref{fig:degree_change} illustrates the relative effect of koreenceine removal on increases vs decreases in vertex degree of the transcriptional orthology correlation network for each species.  In the figure the transcriptional orthology IDs are sorted according to the difference between mutant vs wildtype vertex degree. The blue curve shows the resultant vertex degree difference and the orange curve shows the vertex mean difference. Observe that the order of decreasing differences of vertex degree does not correspond to the order of decreasing differences in vertex mean. However, a change in the vertex mean almost always accompanies a change in vertex degree, although the converse is not true. 
%Thus translates to that there are critical changes in second order interactions which do not appear in first order statistics. This is a complementary information that our model provides, which is useful in analyzing the treatment effects. 
Also note from the asymmetry of the blue curves in Fig. \ref{fig:degree_change} that the mutant's networks have many more vertices that increase than decrease in vertex degree as compared to the wildtype.  Thus koreenceine removal seems to increase network centrality of a large number of transcriptional orthologs, especially for {\it F. johnsoniae}.
We point out that the large spikes that appear in the orange curves (vertex mean difference) for {\it F. johnsoniae} and {\it P. koreensis}, correspond to the ID KXXXXX, which are genes that were not mapped to any Kegg  transcriptional ortholog. Further discussion can be found in the supplementary.
%the mean and degrees of these genomes get effected significantly for P. koreensis, which was also observed in \cite{hurley2022}.  

In summary, the  proposed model can provide two important data analysis components for microbiome model community analysis. First, we can assess transcriptional orthology composition changes under the treatment by observing the means of the marginal distributions provided by the proposed model. Second, we can assess the second order interaction changes by using the correlation networks that are obtained from the covariance matrices of the marginal distributions. These two components along with the abundance ratio analysis in \cite{hurley2022} provide a complementary analysis of microbial model communities, which can further be interpreted by microbiologists.

\section{Conclusion}

A hierarchical Bayesian latent variable model was proposed for the joint analysis of multiple discrete datasets. We explained the associations between the features of the datasets with a common lower dimensional latent space, represented by a set of independent identically distributed Gaussian random variables. To overcome the lack of conjugacy between the multinomial observation distribution and the Gaussiam latent space distribution, we developed a variational EM algorithm based on quadratic bound approximations for estimating the parameters in the model. The computation of the algorithm scales linearly with the number of features, samples, and datasets. Simulation studies show that the proposed model can recover low-rank covariance structures accurately. Furthermore, our real-world microbiome experiment demonstrates  the potential real-world utility of the model for exploration of correlation and associated networks for dichotomous microbiome data. % We highlight several correlations discovered by the proposed model using the abundance dataset of Microbiome communities. 

There are several promising directions for future work. %Here the observations have been assumed to be conditionally independent and identically distributed, which is one limitation of the model. 
One possible area of future work is to incorporate system dynamics into the latent space so as to explicitly capture temporal correlations. In particular, there is increasing interest in collecting longitudinal microbiome data for studying adaptation, resilience, and dynamics over time.  
%provide temporal information attached with the samples taken from the sites. 
Incorporation of a state-space dynamical model into our framework can reveal temporal evolution of the interactions between the genomes. 
%Another limitation of the proposed model is that the induced precision matrix is not enforced to be sparse. 
Another future direction is to improve the parsimony of the model by incorporating sparsity into the latent representation by using sparsity inducing priors for the covariance or inverse covariance (precision) matrices.  
%This would  could be useful for variable selection and for visualization of dominant partial correlations. %Graphical Lasso can enforce this property. 
%However, it is not straightforward for the proposed model to incorporate sparse penalties for covariance matrices due to underlying low-rank decomposition. 

\bibliographystyle{plain}
\bibliography{main.bib}

\begin{appendices}
\section{Estimation of Posterior Parameters} \label{ap:post}
\noindent Log-likelihood of Multivariate Normal distribution $\log \mathcal{N}(\bm{x}|\bm{\mu}, \bm{\Sigma})$ can be written as:
\begin{equation*}
    -\frac{1}{2}\bm{x}^T\bm{\Sigma}^{-1}\bm{x} + \bm{x}^T\bm{\Sigma}^{-1}\bm{\mu} + const
\end{equation*}
in which the second order term in $\bm{x}$ corresponds to the inverse of covariance matrix $\bm{\Sigma}$, and the linear term corresponds to the mean when multiplied with $\bm{\Sigma}$. Inferring the mean and covariance from linear and quadratic terms is called completing the square approach. We make use of this method to infer the posterior distributions of $\bm{z}_{k,i}$, which is denoted as $q(\bm{z}_{k,i} | \bm{m}_{k,i}, \bm{S}_{k,i})$. Given the joint likelihood in Eq. \ref{eq:joint} and quadratic approximation in Eq. \ref{eq:lse_app}, one can collect the quadratic terms in $\bm{z}_{k,i}$ as follows:
\begin{equation*}
    -\frac{1}{2}\bm{z}_{k,i}^T\bm{\Sigma}_k^{-1}\bm{z}_{k,i} - \sum_{l=1}^L \frac{N_{kl,i}}{2} \bm{z}_{k,i}^T \Theta_{kl}^T \bm{A}_{l} \Theta_{kl} \bm{z}_{k,i},
\end{equation*}
which follows Eq. \ref{eq:post_cov} for posterior covariance $\bm{S}_{k,i}$ estimate. Similarly, the linear terms are collected as:
\begin{equation*}
    \bm{\Sigma}_k^{-1} \bm{\mu}_k \bm{z}_{k,i} + \sum_{l=1}^L \bm{x}_{kl,i}\bm{\Theta}_{kl}\bm{z}_{k,i} + N_{kl,i}\bm{b}_{kl,i}^T\bm{\Theta}_{kl}\bm{z}_{k,i}.
\end{equation*}
Collecting the terms and multiplying with the posterior covariance estimate yields posterior mean $\bm{m}_{k,i}$ estimate as given in Eq. \ref{eq:post_mean}.

\end{appendices}

\let\oldsection\section
\renewcommand\section{\clearpage\oldsection}
\section{Supplementary}

\begin{figure}[ht]
    \centering
    \begin{subfigure}{0.50\textwidth}
        \centering
        \includegraphics[width=1\textwidth]{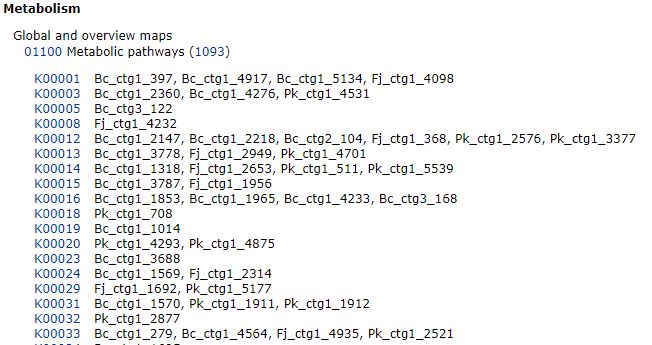}  
        \caption{Functional orthologies of THOR genomes}
    \end{subfigure}\hfill
    \begin{subfigure}{0.50\textwidth}
        \centering
        \includegraphics[width=1\textwidth]{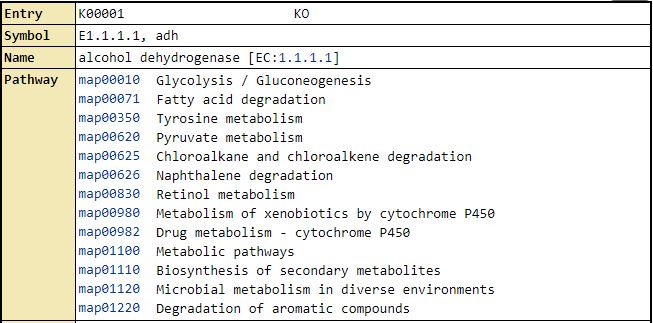} 
        \caption{Kegg Orthology K00001, an alcohol dehydrogenase}
    \end{subfigure}
    \caption{ Examples of classification of THOR transcripts based on metabolic functions inferred from gene sequences. Multiple species may have the same transcriptional orthology. For instance, the Kegg functional orthology
    K00001 is a group of transcipts that encode alcohol dehydrogenase (AD), an enzyme that plays a role in several cellular processes. AD homologues are found in genomes of
    %a metobolic pathway which plays role in multiple core processes, includes genomes from 
    both {\it Bacillus cereus} and {\it Flavobacterium johnsoniae}.}
    \label{fig:orthology}
\end{figure}

\begin{figure*}
    \centering
    \begin{subfigure}{0.33\textwidth}
        \centering
        \includegraphics[width=1\textwidth]{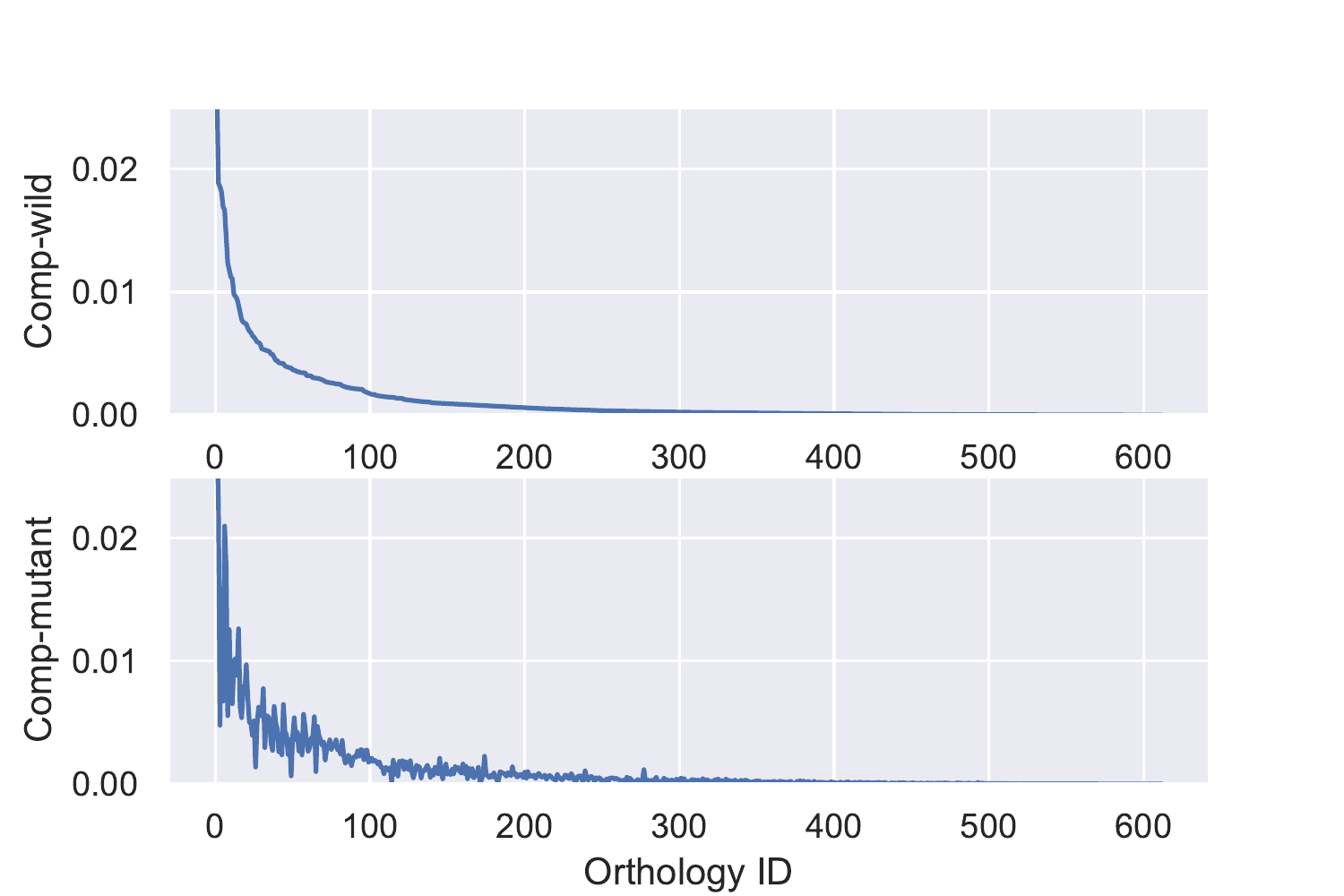}  
        \caption{B. cereus}
    \end{subfigure}\hfill
    \begin{subfigure}{0.33\textwidth}
        \centering
        \includegraphics[width=1\textwidth]{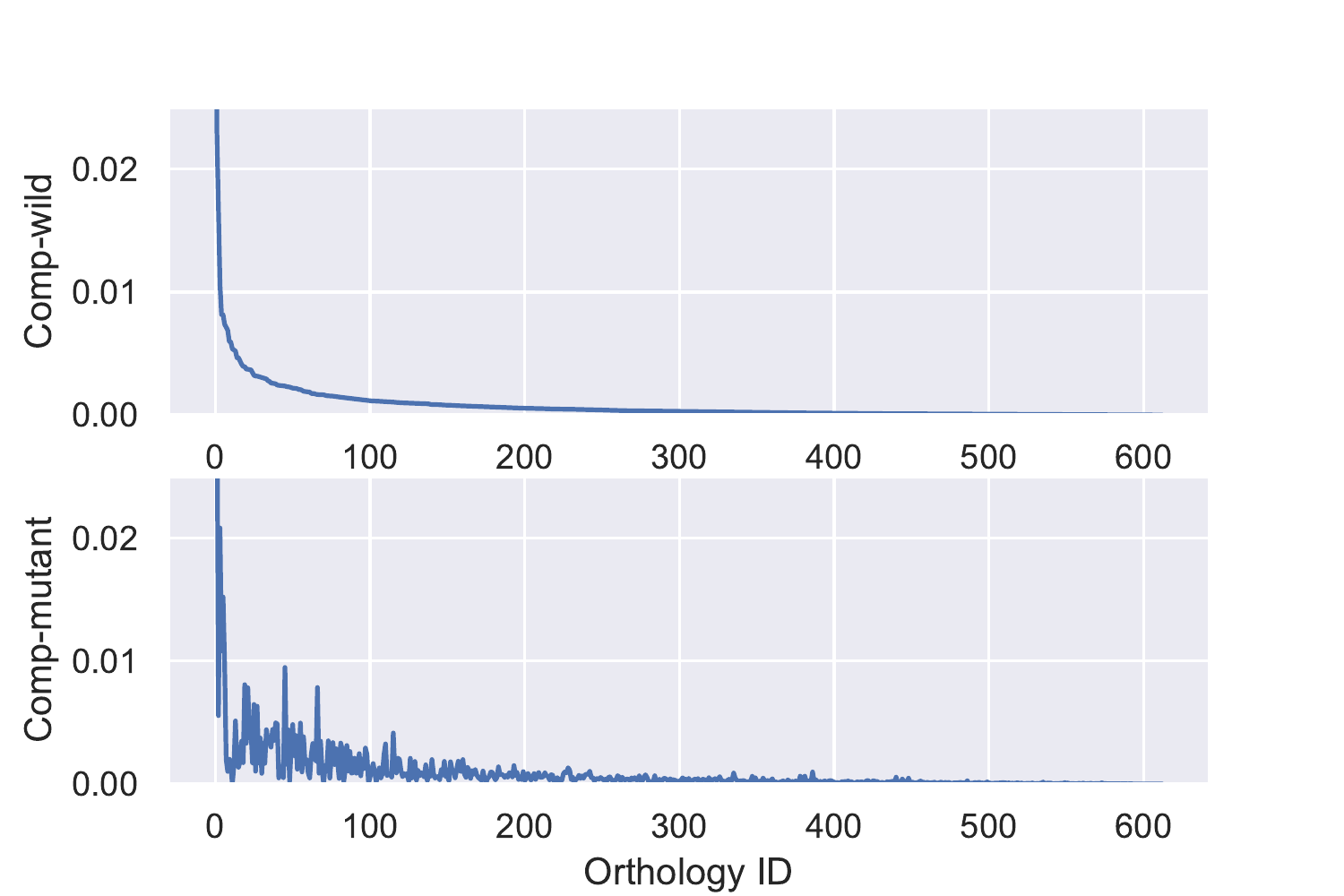} 
        \caption{F. johnsoniae}
    \end{subfigure}
    \begin{subfigure}{0.33\textwidth}
        \centering
        \includegraphics[width=1\textwidth]{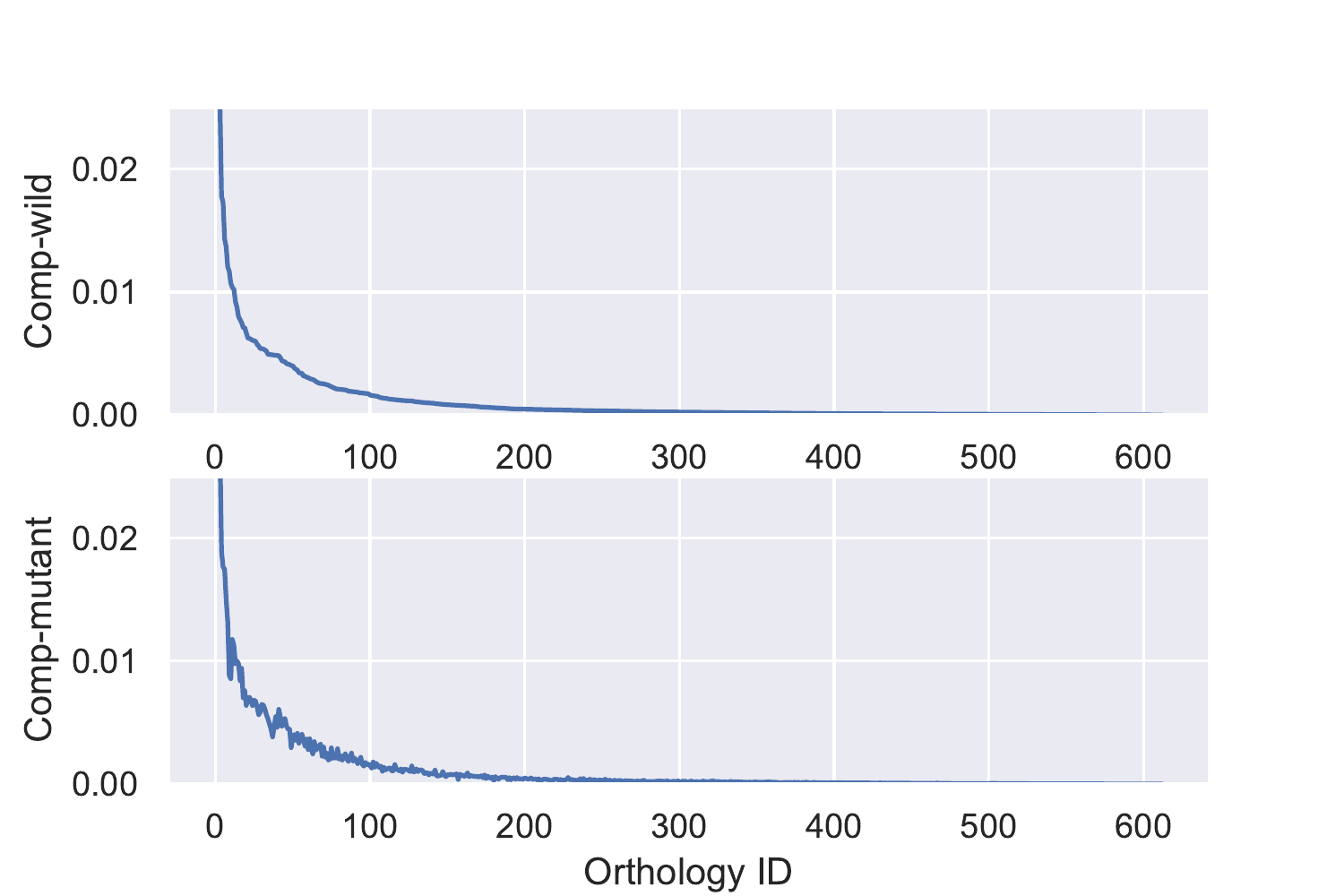} 
        \caption{P. koreensis}
    \end{subfigure}
    \caption{
    %\aoh{I don't understand what you are plotting here - what is "Ratio?" Why don't you simply show barplots of the counts across all the ortholog IDs? This would be more interpretable and easier to compare to Fig 11.}
    Comparison of the effect of koreenceine removal on the mean transcriptional orthology compositions for each of the three species. The ortholog IDs are sorted in decreasing order of the wildtype's Hellinger distance to the mutant composition distribution. Top panel corresponds to wild-type compositions whereas bottom panel shows the mutant compositions, in which the koreenceine pathway of {\it P. koreensis} has been removed, with aligned indexes. Hellinger distances between the estimated probability vectors of the treatments are 0.12, 0.18, 0.05 for {\it B. cereus, F. johnsoniae}, and {\it P. koreensis}, respectively, which suggests that the transcriptional orthology composition of {\it F. johnsoniae} is most affected by koreenceine removal.}
    \label{fig:mean_hist}
\end{figure*}

% Mean change between conditions
\noindent{\bf Mean changes over mutant/wildtype conditions}.  %\aoh{Mehmet: 1) Please clarify whether this section is related at all to the proposed model. My impression is that it is not since I did not see any use of the model parameters in your descriptions of the analysis. If true then we should eliminate this section in addition to Fig. 9, and then we can explain the mean vertex counts in a sentence or two for the purposes of explaining Fig 11. 2) If we keep this section please verify correctness of my  description below as I don't think it is completely correct since I can't make sense out of Figure 9. } 
For each species, and under either of the wildtype or mutant conditions, we estimate the $613$-dimensional mean vector of Kegg functional orthology counts, as determined by the iterative estimate Eq. (17). %\ref{eq:prior_mean}. %using a marginalized density as given in \ref{eq:incuded_dist}. 
We then transform this estimated mean vector to a probability distribution over IDs using the multivariate softmax transformation, yielding what we call the transcriptional orthology probability distribution that quantifies the composition of orthologs for each species under each condition. 
The change to the species-specific probability distribution caused by the removal of koreenceine is quantified by comparing the wildtype and mutant distributions, shown in Fig. \ref{fig:mean_hist}. 
%we show the histogram of 
%shows the histograms of these vectors for the three species. 
To quantify the effect of koreenceine removal, we evaluate the Hellinger distances between the respective wildtype and mutant transcriptional orthology probability distributions. The distances where found to be 0.12, 0.18, and 0.05 for {\it B. cereus, F. johnsoniae}, and {\it P. koreensis,} respectively. This suggests that the transcriptional orthology composition of {\it F. johnsoniae} is most affected under the mutant condition.  In contrast the transcritpional composition of {\it P. koreensis} changes the least of the three species. 

We also observed that the removal of koreenceine had a substantial effect on the relative abundance of different microbial species. For wildtype and mutant conditions the relative abundances of different species in the model community were computed by taking the total number of RNA-Seq counts for each of the 3 species and normalizing it by its sum. The change in relative abundances was quantified by relative change abundance ratio, which are found to be -0.01 (-1\%) for {\it B. cereus}, 0.10 (10\%) for {\it F. johnsoniae}, and -0.08(-8\%) for {\it P. koreensis}, respectively. These quantities are compatible with \cite{hurley2022}, which suggests that deleting the koreenceine biosynthetic pathway in  {\it P. koreensis} enhances growth of {\it F. johnsoniae} and does not affect {\it B. cereus}.
%was unaffected by the loss of koreenceine.
%and  the growth rate of {\it P. koreensis} was enhanced. 
It is interesting that while the  removal of koreenceine does not affect the abundance of {\it B. cereus} it has substantial effect on the transcriptional orthology probability distribution, as predicted by our model,  discussed in the previous paragraph. %change in different order than the relative abundance ratios, i.e., 
%show that the {\it B. cereus} orthology composition was more affected by loss of koreenciene than was {\it  P. koreensis}. 

\begin{table*}[ht]
    \small
    \centering
    \begin{tabular}{lcccccc}
    \hline
        \textbf{Orthology ID} & \textbf{DegDiff} & \textbf{DegW} & \textbf{DegM} & \textbf{MeanDiff} & \textbf{MeanW} & \textbf{MeanM} \\
        \hline
        \textbf{\textit{Emerging Connections}} \\
        \hline
        K15777 (4,5-DOPA dioxygenase extradiol) &	\textbf{275} &	1   &	276 &	-0.08     &	0.10 &	0.01\\
        K07646 (OmpR family, sensor histidine kinase KdpD) &	\textbf{229} &	94  &	323 &   0.00    &	0.00 &	0.00\\
        K03040 (DNA-directed RNA polymerase subunit alpha) &	\textbf{227} &	32  &	259 &   6.16    &	1.68 &	7.84\\
        \hline
        \textbf{\textit{Extinguished Connections}} \\
        \hline
        K01610 (phosphoenolpyruvate carboxykinase (ATP)) &	\textbf{-132} &	135 &	3 &	-3.29 &	4.64 &	1.35\\
        K01358 (ATP-dependent Clp protease, protease subunit) &	\textbf{-136} &	137 &	1 &	-5.88 &	6.89 &	1.01\\
        K00274 (monoamine oxidase) &	\textbf{-184} &	186 &	2 &	-0.00 &	0.00 &	0.00\\
        \hline
        \textbf{\textit{Up-regulated Functions}} \\
        \hline
        K02358 (elongation factor Tu) &	104 &	225 &	329 &	\textbf{11.57}  &	23.02 &	34.59\\
        K02355 (Not found in KEGG) &	104 &	189 &	293 &	\textbf{10.22} &	10.59 &	20.81\\
        K03076 (preprotein translocase subunit SecY) &	165 &	78  &	243 &	\textbf{7.12} &	2.35  &	9.48\\
        \hline
        \textbf{\textit{Down-regulated Functions}} \\
        \hline
        KXXXXX (Unannotated) &	139 &	207 &	346 &	\textbf{-57.08} &	536.04 &	478.95\\
        K04043 (molecular chaperone DnaK) &	30  &	212 &	242 &	\textbf{-10.73} &	16.28 &	5.55\\
        K01358 (ATP-dependent Clp protease, protease subunit) &	-136 &	137 &	1   &	\textbf{-5.88} &	6.89 &	1.01\\
        \hline
        \textbf{\textit{Emerging Connections w/o regulations}} \\
        \hline
        K07646 (OmpR family, sensor histidine kinase KdpD) &	\textbf{229} &	94  &	323 &	\textbf{0.00} &	0.00 &	0.00\\
        K02238 (Not found in KEGG) &	\textbf{161} &	151 &	312 &	\textbf{0.00} &	0.00 &	0.00\\
        K00788 (thiamine-phosphate pyrophosphorylase) &	\textbf{129} &	177 &	306 &	\textbf{0.00} &	0.00 &	0.00\\
        \hline
        \textbf{\textit{Extinguished Connections w/o regulations}} \\
        \hline
        K00274 (monoamine oxidase) &	\textbf{-184} &	186 &	2  &	\textbf{-0.00} &	0.00 &	0.00\\
        K03781 (catalase) &	\textbf{-109} &	175 &	66 &	\textbf{0.00} &	0.00 &	0.00\\
        K02575 (MFS transporter, nitrate/nitrite transporter) &	\textbf{-50}  &	150 &	100 &	\textbf{0.00} &	0.00 &	0.01\\
        \hline
        \textbf{\textit{Up-regulated Functions w/o degree changes}} \\
        \hline
        K02495 (oxygen-independent coproporphyrinogen III oxidase) &	\textbf{0} &	1 &	1 &	\textbf{0.81} &	0.16&	0.98\\
        K03530 (Not found in KEGG) &	\textbf{0} &	1 &	1 &	\textbf{0.67} &	0.22&	0.89\\
        K02899 (large subunit ribosomal protein) L27 &	\textbf{1} &	1 &	2 &	\textbf{0.64} &	0.91&	1.56\\
        \hline
        \textbf{\textit{Down-regulated Functions w/o degree changes}} \\
        \hline
        K03695 (ATP-dep Clp protease ATP-binding subunit ClpB) &	\textbf{-2} &	3 &	1 &	\textbf{-1.37} &	1.86 &	0.48 \\
        K00265 (glutamate synthase (NADPH) large chain) &	\textbf{0}  &	1 &	1 &	\textbf{-0.73} &	0.96 &	0.22\\
        K03733 (Not found in KEGG) &	\textbf{0}  &	1 &	1 &	\textbf{-0.65} &	0.91 &	0.25\\
        \hline
    \end{tabular}
    \caption{ Top 3 transcripts of {\it Flavobacterium johnsoniae} that change the most according to the eight different treatment effect definitions (emerging connection, extinguished connections, up-regulated functions, down-regulated functions, emerging connections w/o regulations, extinguished connections w/o regulations) . Note that the numbers in the 3 last columns are scaled by $10^3$ to reduce clutter in rendering the table. This table provides a snapshot of the transcriptional orthologs in {\it Flavobacterium johnsoniae} that are the most sensitive to removal of koreenceine under 8 different sensitivity criteria.  We highlight {\it Flavobacterium johnsoniae} here because it is the species that is most affected although analogous CSV tables for other species can be found in the supplementary files. Emerging and extinguished connections correspond to, respectively, positive and negative centrality changes, as measured by vertex degree, in the transcriptional ortholog correlation graph. Up-regulated and down regulated functions correspond to the orthologies whose compositions change positively and negatively, respectively. Emerging and extinguished connections w/o mean change denote transcriptional orthologs for which there are changes in vertex degree but no change in vertex mean. Up-regulated and down regulated w/o degree changes denote transcriptional orthologs that change have changes in mean but not in vertex degree. The  proposed model can provide two important data analysis components for microbiome model community analysis. First, we can assess transcriptional orthology composition changes under the treatment by observing the means of the marginal distributions provided by the proposed model. Second, we can assess the second order interaction changes by using the correlation networks that are obtained from the covariance matrices of the marginal distributions. These two components  provide a complementary analysis of microbial model communities, which can further be interpreted by microbiologists. }
    \label{tab:top_3}
\end{table*}

\end{document}